\documentclass[11pt]{article}
\usepackage[a4paper,margin=19mm]{geometry}
\usepackage{amsmath,amssymb,mathtools,booktabs,graphicx,float,microtype}
\usepackage[dvipsnames]{xcolor}
\usepackage{wasysym}
\usepackage{jheppub}

\newcommand{\A}{\mathcal A}
\newcommand{\mP}{\mathcal P}
\newcommand{\cZ}{\mathcal Z}
\newcommand{\cP}{\mathtt P}
\newcommand{\E}{\mathbb E}
\newcommand{\Prob}{\mathbb P}
\newcommand{\dd}{\mathrm d}
\newcommand{\bz}{\boldsymbol z}
\newcommand{\ba}{\boldsymbol a}

\newcommand{\one}{\mathbf 1}
\newcommand{\1}{\mathbf 1}
\newcommand{\ndt}{\noindent}

\title{\textbf{The origin of amplitude chaos and strings}}
\author[\textsuperscript{\(\smiley\)}]{Pallab Basu,}
\author[\textsuperscript{\(\blacksmiley\)}]{Anurag Sarkar}

\affiliation[\textsuperscript{\(\smiley\)}]{Mandelstam Institute for Theoretical Physics, School of Physics,\\ University of the Witwatersrand, Johannesburg, Wits 2050, South Africa.}
\affiliation[\textsuperscript{\(\blacksmiley\)}]{National Institute for Theoretical and Computational Sciences, \\ School of Physics, and Mandelstam Institute for Theoretical Physics,\\ University of the Witwatersrand, Johannesburg, Wits 2050, South Africa.}

\emailAdd{pallab.basu@wits.ac.za}
\emailAdd{anurag.sarkar@wits.ac.za}

\abstract{
We analytically study the peak statistics of an amplitude ensemble to identify the local mechanism
responsible for peak repulsion.  Our analytical approaches show that a smooth, nondegenerate ensemble of sufficiently differentiable amplitudes
exhibits quartic nearest-neighbor repulsion, \(p_{\rm nn}(S)\sim S^4\), where $S$ is the spacing between adjacent peaks.  We verify this prediction
numerically for the tree-level highly excited string scattering
amplitudes.  Interestingly, we find the quartic repulsion to be robust by varying both the underlying ensemble and the amplitude functions,
whereas a tunable cusp model shows that weakening differentiability changes the strength of the repulsion.
Peak repulsion therefore defines a local universality class that is
not unique to string theory and may occur in amplitude ensembles in other theories.
}

\begin{document}
	\maketitle


\section{Motivation and scope}

In a closed Hamiltonian system, quantum chaos is commonly diagnosed through the statistics of
energy levels~\cite{BohigasGiannoniSchmit1984,GuhrEtAl1998,Haake2018}. An especially clean indicator of quantum chaos is the repulsion between nearest neighbour eigenvalues, which is denoted as the level-repulsion. Scattering problems have no discrete energy spectrum of that kind, so one instead
looks for statistical structure in the \(S\)-matrix: resonances, time delays, or the response of an
exclusive amplitude to small changes of kinematics or of the incoming state
~\cite{OttTel1993,FyodorovSommers1997,Rosenhaus2021}. As indicated by the theoretical~\cite{Smilansky1991,BluemelSmilansky1988,BluemelSmilansky1990,MitchellEtAl2010} and experimental~\cite{DoronSmilansky1990,KudrolliEtAl1994,StoffregenEtAl1995,HemmadyEtAl2005} studies, Bl\"umel and Smilansky~\cite{BluemelSmilansky1988,BluemelSmilansky1990} hypothesized: \emph{the statistical properties of the S-matrix are determined by the Random-matrix theory (RMT)}. Although RMT provides an important benchmark for
these observables, but agreement with an RMT statistic is a diagnostic of chaos, not by
itself an explanation of the underlying mechanism.

Highly excited strings make this distinction sharp.  At a fixed mass level \(N\), the
large degeneracy of string states supplies an ensemble of microscopic preparations, while their
tree-level decay and scattering amplitudes remain explicitly computable.  The angular dependence
of these amplitudes is highly sensitive to the oscillator content of the chosen state
~\cite{GrossRosenhaus2021}.  This sensitivity has been studied through peak spacings and related
RMT-inspired observables~\cite{BianchiEtAl2022,BianchiEtAl2023,HashimotoEtAl2022,Firrotta2023,DasMandalSarkar2024,BianchiEtAl2026}.  In these examples, it was found that the nearest neighbouring peaks show peak-repulsion, indicating towards the RMT-inspired chaotic scattering.  Yet the
available results do not reduce to one clean RMT ensemble~\cite{SavicCubrovic2024,Pesando2026}: the three-point amplitude with one
highly excited string and two tachyons (HTT) does not saturate to a Gaussian RMT spacing distribution for large $N$~\cite{BianchiEtAl2023}, whereas
ensembles with two highly excited strings and one tachyon (HHT) can display an approximately
GOE-like bulk distribution in particular sectors~\cite{DasMandalSarkar2024}.  What is common in all of these amplitudes is the suppression of the probability of very
small spacings between neighboring angular peaks, i.e., the peak-repulsion.

This repulsion poses a more basic question than the quality of a global RMT fit.  In an RMT, the repulsion is observed between the eigenvalues of a matrix, whose entries are drawn prom a probability distribution; hence the process of chossing the matrix is effectively random. The eigenvalue repulsion in RMT follows from the Jacobian that appears when matrix entries are replaced by
eigenvalues and eigenvectors~\cite{Forrester2010}.  But the peaks of a string scattering amplitude are not eigenvalues: they are local
extrema of an exact function of the scattering angle, and no matrix is diagonalized to obtain
them.  Moreover, the amplitude ensemble is generated by sampling the partitions of the string excitation level, which are exactly known~\cite{GrossRosenhaus2021,BianchiEtAl2023}. Hence there is no ``randomness'' in this amplitude ensemble.  The RMT analogy therefore does not explain why two amplitude peaks avoid contact.  Hence we ask: \emph{what local event is required for two physical peaks to approach one another, and
what probability does an amplitude ensemble assign to that event?}

\subsection{Central questions and main results}

    In this work, we address the following questions:
    \begin{itemize}
        \item How can we explain the peak repulsion of the string amplitudes analytically?
          \begin{center}
\fbox{\begin{minipage}{0.90\textwidth}
\textbf{\large Main result I: the string-amplitude contact law.}

\smallskip
\noindent
The neighboring peaks of the highly excited string amplitude exhibit \emph{quartic contact
repulsion},
\[
p_{\rm nn}(S)\sim C S^4,\qquad S\downarrow0,
\]
where $S$ denotes the peak spacing. This power is derived analytically in two equivalent ways:  a direct calculation of the collision of two neighboring physical peaks (see \eqref{eq:direct LS 1 compact}), and an equivalent description regarding three consecutive roots (see \eqref{eq:Wdefinition}). This is supported by the finite but large level string amplitude numerics.
\end{minipage}}
\end{center}
        \item Is the peak repulsion in the amplitudes special to the excited strings (i.e, if it depends sensitively on the amplitude functions and the nature of microstate ensemble), or can they be observed in a generic ensemble of analytic amplitudes?   What controls the strength of the peak repulsion?

        \begin{center}
\fbox{\begin{minipage}{0.90\textwidth}
\textbf{\large Main result II: a universality class beyond strings.}

\smallskip
\noindent
Quartic contact repulsion is \emph{not uniquely string-theoretic}.   We check this explicitly in Section~\ref{sec:examples}: by varying both the ensemble and the smooth amplitude functions, all of which support this universality. There we also show
that reduced differentiability can produce sub-quartic exponents.  The same local repulsion can
therefore occur in suitable amplitude ensembles in other theories.
\end{minipage}}
\end{center}
      
    \end{itemize}

    The string example we will look into is the three-point amplitude with one highly excited string and two tachyons (HTT), which can be expressed as a function of the scattering angle $\alpha$.  To state the problem precisely, we regard an amplitude \(\A(\alpha)\) as a real function of one angular variable.  A \emph{physical peak} is a local maximum of \(|\A|\); it may be either a positive maximum or a negative minimum of \(\A\). Then the nearest-neighbor peak spacing distribution $p_{\rm nn}(S)$ and the contact exponent $\beta$ are defined by
    \[
	p_{\rm nn}(S)\sim C S^\beta,
	\qquad S\downarrow0.
    \]

    We answer the analytic questions in two complementary ways:  a direct calculation of the collision of two neighboring physical peaks in Section~\ref{sec:spacing analysis}, and an equivalent description regarding three consecutive roots in Section~\ref{sec:roots analysis}.  Both these approaches result into a quartic peak repulsion in the small spacing limit: $p_{\rm nn}(S)\sim S^4$.  These analysis indicate that the collection of the peaks in an amplitude ensemble can be viewed as a random process, where the amplitude derivatives are being sampled from a continuous distribution.  In the consecutive roots analysis, we find an explicit Vandermonde factor which results into the peak repulsion. Note that a similar Vandermonde factor is responsible for level-repulsion in the RMT, although its origin there is different. In the following Section \ref{sec:HTTnumerics}, we carefully compute the small peak-spacing distribution of the HTT amplitudes, which we find to be consistent with the quartic repulsion and the corresponding analytic descriptions.

    The second result is broader.  Section~\ref{sec:examples} replaces the HTT formula and its partition ensemble by analytic partition-sine amplitudes and by random Fourier amplitudes with
several smooth coefficient laws.  These ensembles again exhibit the quartic contact law.  Thus
the exponent does not rely on the detailed string functions or on a finely tuned distribution of
string microstates.  Analyticity alone is not the complete assumption: the ensemble must also
sample the relevant local derivative data nondegenerately and with finite, nonzero density.
Within this regular class, however, quartic contact repulsion is generic and should therefore be
expected in other real amplitude ensembles, including those arising in theories unrelated to
strings.

Finally, a tunable cusp family shows what happens outside this class.  Reducing the local
differentiability changes the contact exponent continuously and produces sub-quartic repulsion.
The combined analytic and numerical evidence therefore identifies both a universality class and
its boundary.


	\section{The spacing behaviour of two adjacent peaks}
	\label{sec:spacing analysis}

    As explained in the introduction, our main focus of this work will be based on the three point amplitude between one highly excited string at level $N$ and two tachyons. This amplitude can be expressed as a function of an angular scattering parameter $\alpha$~\cite{GrossRosenhaus2021,BianchiEtAl2022}. The ensemble of amplitudes are generated from the string microstates, i.e., the integer partitions of $N$. For each microstate, the amplitude can be explicitly written as an exact analytic function.

    This setup therefore motivates us to consider an ensemble $\{ \A(\alpha) \}$, consisting of generic functions $\A(\alpha)$, which we denote as ``amplitudes''. Each function $\A(\alpha)$ in the ensemble is exactly known; there is no notion of randomness (for example, unlike the functions appearing in a Brownian motion scenario~\cite{CramerLeadbetter1967,AdlerTaylor2007}). We will assume that these functions are well behaved, or at least the first three derivatives are continuous (hence the functions are at least $C^3$). 
    
	Then let \(\A(\alpha)\) be a real signed amplitude in this ensemble. The peaks of this amplitude are denoted by the positions of the local maxima of $|\A(\alpha)|$. Therefore the peak process is
	\begin{equation}
		\mathtt{P} = \{\alpha: \A'(\alpha)=0,\ \A(\alpha) \A''(\alpha)<0\}.
		\label{eq:peakcondition}
	\end{equation}
	
	Now, let two very nearby adjacent peaks be located at $\alpha = \pm h$. From the string amplitude structure, we see that for two adjacent peaks $\A$ crosses zero in between them (see Fig.~\ref{fig:simplehttpeaks}). This implies, one of them is a local minima with negative value and the other one is a local maxima with a positive value. Hence the conditions obeyed by the two adjacent peaks are:
	\begin{align}
			\A'(\pm h)=0,\qquad
			\A(\pm h)\A''(\pm h)<0,\qquad
			\A(-h)\A(h)<0.
			\label{eq:2peaksCond}
		\end{align}

	Since we are interested in small spacing behaviour, let us consider $h$ to be small. Then, we can work with the Taylor expansion:
	\begin{align}
		\A(x) = \A(0) + \A'(0) x + \frac{\A''(0)}{2}x^2
		+\frac{\A'''(0)}{6}x^3+O(x^4).
		\label{eq:taylor}
	\end{align}
	With this exapansion, we can now use the peak conditions in Eq.\eqref{eq:2peaksCond} to get the following:
	\begin{itemize}
		\item 
		 From the first condition $\A' (\pm h) = 0$, one obtains:
		 \begin{align}
		 	a_1=-\frac{h^2}{2}a_3+O(h^3),\qquad a_2=O(h^2),
		 	\label{eq:a1 a2 leading}
		 \end{align}
		 where, $a_k \equiv \A^{(k)} (0)$. Thus, the first and second derivatives vanish with $h \to 0$. So \(a_3\) fixes the local shape of \( \A'\) and hence the way the two solutions of \(\A'=0\) come together.  Now we can approximate,
		 \begin{align}
		 	A(x)=a_0+ a_3 \left(\frac{x^3}{6}-\frac{h^2x}{2}\right)+O(h^4),
		 	\qquad \A'(x)=\frac{a_3}{2}(x^2-h^2)+O(h^3),
		 	\qquad x=O(h).
		 	\label{eq:a3 range}
		 \end{align}
		 
		 \item
		 The last condition in Eq.\eqref{eq:2peaksCond}, along with the above approximation then implies,
		 \begin{align}
		 	- \frac{|a_3| h^3}{3} < a_0 < \frac{|a_3| h^3}{3}.
		 \end{align}
		 Then, $a_0 \to 0$ when $h \to 0$.
	\end{itemize}

	So we see that near two adjacent peaks, the amplitude is controlled by the first four coefficients of the Taylor series: $(\A(0),\A'(0),\A''(0),\A'''(0))$, in the leading order. The above conditions together also then imply:
	\begin{align}
		(\A(0),\A'(0),\A''(0),\A'''(0))  \xrightarrow{ h \to 0 } (0,0,0,c),
        \label{eq:coeffs limit}
	\end{align}
	where $c$ is some non-zero finite number. With these conditions, we now look into an ensemble of such amplitudes.

	\subsection{Ensemble average and the small spacing behaviour}

    Our setup considers an ensemble of explicit functions, so for each element of this ensemble the amplitude is deterministic. For each of these amplitudes, we have a set of adjacent peak pairs. Now, from our discussion in the previous section we see that the behaviour of the adjacent peaks are completely determined in terms of the first three derivatives of the amplitude. We express the four leading coefficients as
	\begin{equation}
		J=(\A(0),\A'(0),\A''(0),\A'''(0)).
	\end{equation}
    
    Then all the adjacent peak pairs of the ensemble $\{\A\}$ together effectively generate a joint probability distribution of these four coefficients, i.e., sampling the amplitudes results into the four-component vector \(J\) is being drawn from a particular distribution.  We assume
	that it has a continuous, full-dimensional probability density $p_J(j_0,j_1,j_2,j_3)$. The probability for the four quantities to lie in a small box of side lengths \(\dd j_0,\ldots,\dd j_3\) is,
	\begin{equation}
		\mathbb{P} (J\in\dd^4j)=p_J(j_0,j_1,j_2,j_3)\,\dd^4j = p_J(j_0,j_1,j_2,j_3)\,\dd j_0\,\dd j_1\,\dd j_2\,\dd j_3.
		\label{eq:jetdensity}
	\end{equation}
	
	Here, full-dimensional means that the four quantities are not forced to obey an additional exact relation such as \(g(\A,\A',\A'',\A''')=0\).  Equivalently, the probability is spread through an ordinary four-dimensional neighbourhood.  This matters because a two-peak collision restricts the allowed ranges of the coefficients. The product of the widths of those ranges contribute to the small-spacing power.  If the ensemble were concentrated on a lower-dimensional set (i.e., constrained through some exact relation between the components), that power counting could change. Also, for our calculations, we require \(p_J\) to be finite, continuous, and nonzero for these generic two-peak configurations. Then following \eqref{eq:coeffs limit}, for vanishing gap $h \to 0$ we can smoothly approximate,
	\begin{align}
		p_J(j_0,j_1,j_2,j_3) \xrightarrow{h \to 0} p_J(0,0,0,c).
		\label{eq:pJ approx}
	\end{align}
	
	\subsubsection{One-point and two-point density}
	
	Let $\{ \alpha_i \}$ be the solutions of the equation $\A' (\alpha) = 0$. Then the factor $\sum_i \delta (\alpha - \alpha_i)$ counts the number of all the solutions in a region when it is integrated over $\alpha$. From these, we need to pick the solutions which also satisfy $\A (\alpha_i) \A'' (\alpha_i) < 0$. Hence the local peak counting density can be expressed as~\cite{Kac1943,Rice1944,CramerLeadbetter1967,AdlerTaylor2007,AzaisWschebor2009},
	\begin{align}
		\rho_{\mathtt{P}} = \sum_i \delta (\alpha - \alpha_i) \, \one_{\A(\alpha)\A''(\alpha)<0},
	\end{align}
	where $\one_{\A(\alpha)\A''(\alpha)<0}$ is an operator that chooses only the solutions satisfying $\A (\alpha_i) \A'' (\alpha_i) < 0$. So this factor can be effectively expressed as the Heaviside step function $ \Theta ( - \A(\alpha)\A''(\alpha) ) $. For the simple zeros of \( \A'\), with the identity
	\begin{equation*}
		\delta(\A'(\alpha)) \, |\A''(\alpha)|
		=\sum_{\A'(\alpha_i)=0}\delta(\alpha-\alpha_i),
	\end{equation*}
	we can rewrite the peak density function as,
	\begin{equation}
		\rho_{\mathtt{P}}(\alpha) = \delta(\A'(\alpha))|\A''(\alpha)|
		\one_{\A(\alpha)\A''(\alpha)<0}.
	\end{equation}
	
	Now we are ready to define the adjacent two-peak density, where the two peaks are located at $\alpha = \pm s/2$ with respect to some reference point $\alpha = 0$, so the peaks are located with a gap $s$. Then the two-point function is simply the expectation value of having two peaks at $\pm s/2$:
	\begin{equation}
		\begin{aligned}
			\mP_{\cP}(s)
			&= \E \left[ \,  \rho_{\cP}(-s/2) \rho_{\cP}(s/2)  \,   \right] \\
			&= \E \left[ \, \delta(\A'(-s/2))\delta(\A'(s/2)) \;
			|\A''(-s/2) \A''(s/2)| \times\one_{\A \A''(-s/2)<0}
			\one_{\A \A''(s/2)<0}   \,  \right].
		\end{aligned}
		\label{eq:KR}
	\end{equation}
	
	We can now perform the complete integral explicitly, beginning with the joint probability density from \eqref{eq:jetdensity}.  Keeping only the leading terms in \(s\), the pair density is,
	\begin{align}
		\mP_{\cP}(s)
		=\int_{\mathbb R^4}&\dd a_0\dd a_1\dd a_2\dd a_3\,
		p_J(a_0,a_1,a_2,a_3) \notag\\
		&\times\delta\left(a_1-ha_2+\frac{h^2}{2}a_3\right)
		\delta\left(a_1+ha_2+\frac{h^2}{2}a_3\right) \notag\\
		&\times\left|(a_2-ha_3)(a_2+ha_3)\right|
		\one_{\text{both points are peaks of }|\A|},
		\label{eq:fullfourintegral}
	\end{align}
	where $h = s/2$. While writing the above, we have used the Taylor expansion discussed in Eq.\eqref{eq:taylor}. In the leading order in $s$, the solution of the delta fuctions are given in Eq.\eqref{eq:a1 a2 leading}. Hence we obtain, 
	\begin{equation}
		\delta\left(a_1-ha_2+\frac{h^2}{2}a_3\right)
		\delta\left(a_1+ha_2+\frac{h^2}{2}a_3\right)
		=
		\frac1s\,
		\delta\left(a_1+\frac{s^2}{8}a_3\right)\delta(a_2).
	\end{equation}
	
	\ndt Similarly, in the leading order, the factor $|\A''(-s/2) \A''(s/2)|$ becomes,
	\begin{align}
		 \left|(a_2-ha_3)(a_2+ha_3)\right|
		=\frac{s^2a_3^2}{4}.
	\end{align}
	
	\ndt And following Eq.\eqref{eq:a3 range}, the integration range of $a_3$ can be confined to the region,
	\begin{align}
		I_s(a_3)=
		\left(-\frac{|a_3|s^3}{24},\,\frac{|a_3|s^3}{24}\right).
	\end{align}
	
	\ndt Putting all these together, the $a_1, a_2$ integrals can be performed to reduce Eq.\eqref{eq:KR} to:
	\begin{align}
		\mP_{\cP}(s)
		=\frac1s\int_{\mathbb R}\dd a_3\,
		\frac{s^2a_3^2}{4}
		\int_{I_s(a_3)}\dd a_0\,
		p_J\left(a_0,-\frac{s^2a_3}{8},0,a_3\right)
	\end{align}
	
	Since we are considering small $s$ behaviour, then the joint probability density $p_J$ can be smoothly approximated according to Eq.\eqref{eq:pJ approx}, to obtain in the leading order,
	\begin{align}
		\int_{I_s(a_3)}\dd a_0\,
		p_J\left(a_0,-\frac{s^2a_3}{8},0,a_3\right)
		\simeq
		\frac{|c|s^3}{12}\,p_J(0,0,0,c).
	\end{align}
	
	\ndt Hence finally we can approximate the leading behaviour of peak spacing distributio for small $s$,
	\begin{equation}
		\boxed{\mP_\cP (s) \simeq \frac{s^4}{48}\int_{\mathbb R}|c|^3
			p_J(0,0,0,c)\,\dd c}, 
		\label{eq:direct LS 1}
	\end{equation}
	or in a compact way,
	\begin{equation}
		\boxed{ \mP_\cP (s) \sim C s^4. }
        \label{eq:direct LS 1 compact}
	\end{equation}
	Thus, the nearby neighbouring peaks show quartic repulsion for small spacings.

	\subsection{Implications and limitations}
	
	First, we note that the calculations worked out in this section are solely based on two particular assumptions: 
	\begin{enumerate}
		\item the spacing $s$ between two peaks are very small, so essentially we are looking at the colliding limit of two peaks,
		\item in this limit the underlying probability distribution is smooth and finite (i.e., the integration in Eq.\eqref{eq:direct LS 1} converges to a finite non-zero value).
	\end{enumerate}
	
	\ndt The quartic repulsion arises entirely from the analytic structure of the amplitude; the details of the ensemble is contaied in the integrated factor in Eq.\eqref{eq:direct LS 1}, which appears as a multiplicative coefficient. This implies that the quartic repulsion is \emph{not} a consequence of the specifics of the distribution or the specifics of the functions appearing in the amplitude. If one samples \emph{any} analytic function from a well defined smooth distribution, one would obtain this quartic repulsion in the small spacing limit. We consider several toy examples in Section~\ref{sec:examples}, and obtain the same quartic repulsion in the small spacing limit.
	
	Secondly, we emphasize here that the quartic repulsion is the universal property in the small spacing limit; it does not fix the nature of the peak-spacing distribution at moderate and large spacings. In generic Gaussian ensembles (GOE, GUE or GSE), the only non-trivial parameter that appears in the level-spacing distribution is the Dyson index $\beta_D$ (= 1, 2, 4 respectively), and hence the entire level-spacing distribution is fixed in terms of this index: $\mP (s) \sim s^{\beta_D} \exp(a_{\beta_D} s^2)$~\cite{Forrester2010}. But in the case of a generic ensemble of amplitudes, there might be numerous parameters, and the exact peak-spacing distribution might generally depend on these parameters. Thus, our analysis of this section and the subsequent ones solely explore the small spacing universal behaviour, and does not fix the entirity of the spacing distribution. Hence, this quartic repulsion does not contradict the absense of any specific Gaussian ensemble in the large $N$ examples of the HTT amplitudes, or the effective GOE-like behaviour observed in the spacing distribution of diagonal HHT amplitudes~\cite{BianchiEtAl2022,BianchiEtAl2023,DasMandalSarkar2024}.


	\section{Equivalent three-roots description}
	\label{sec:roots analysis}
	
\begin{figure}[h]
		\centering
		\includegraphics[width=0.7\linewidth]{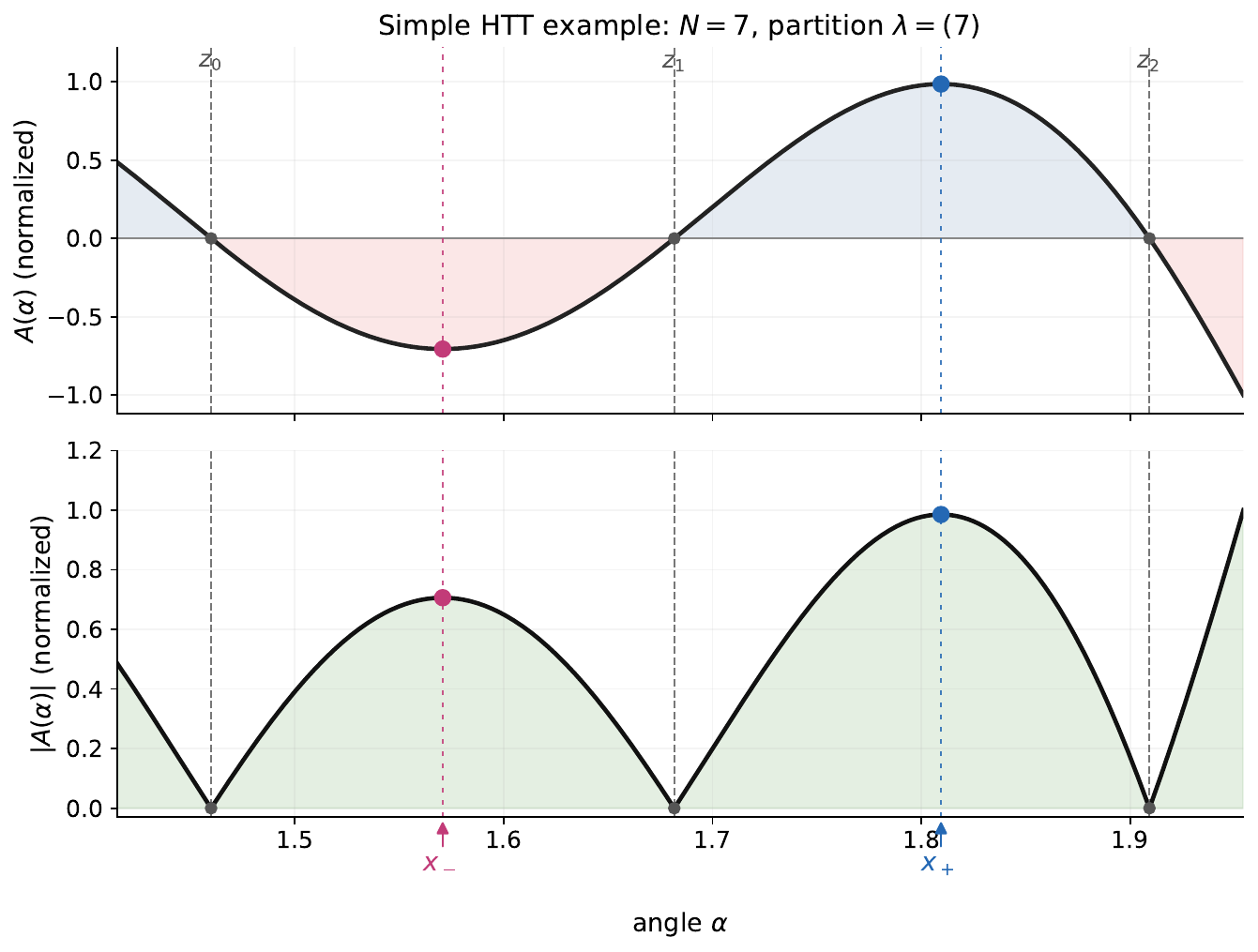}
		\caption{An actual HTT example with $N=7$ and partition $\lambda=(7)$.  The
			upper panel shows the signed amplitude: $x_-$ is a negative dip and $x_+$ is a
			positive peak.  The lower panel shows $|\A|$, for which both $x_-$ and $x_+$ are
			peaks.  The three consecutive roots $z_0,z_1,z_2$ bound the two peak-containing
			cells.  Numerically,
			$z_0=1.4599$, $z_1=1.6817$, $z_2=1.9092$,
			$x_-=1.5708$, and $x_+=1.8094$.}
		\label{fig:simplehttpeaks}
	\end{figure}

	The small-spacing behaviour discussed in the previous section can also be obtained by looking at the coinciding zeros of the amplitude, which we discuss in this section. We denote the zeros of the amplitude $\A (\alpha)$ as roots. Let three consecutive simple roots of an amplitude be located at $\alpha = z_0, z_1, z_2$ with:
	\begin{equation}
		z_0<z_1<z_2,
		\qquad u=z_1-z_0>0,
		\qquad v=z_2-z_1>0.                                  \label{eq:gaps}
	\end{equation}
	It turns out, a peak is located between two consecutive roots. Hence there is one peak of $|\A|$ in the interval $(z_0,z_1)$ and the neighbouring peak is in $(z_1,z_2)$.  Thus a pair of adjacent peaks shares the middle root and is controlled by the root triple $(z_0,z_1,z_2)$. We denote the location of these two peaks with $x_{\pm}$, see Figure~\ref{fig:simplehttpeaks} for example. To analyze the small peak spacing behaviour we need $|x_+-x_-| \to 0$, which implies $u, v \to 0$. In the following discussions, we will focus on this limit.
	
	Since the amplitudes are analytic functions, then in the neighbourhood of these three roots the amplitude can be approximated as:
	\begin{equation}
		\A(\alpha)=B(\alpha)\,(\alpha - z_0)^{m_0} (\alpha - z_1)^{m_1} (\alpha - z_2)^{m_2}  = B(\alpha) Q(\alpha),                              
		\label{eq:localfactor}
	\end{equation}
	where $B$ contains the remaining roots and all nonzero smooth factors, and $\boldsymbol m=(m_0,m_1,m_2)$ are the root multiplicities, so $m_i\geq1$. 
	
	 If $B'/B$ stays finite while $u,v\to0$, the peak locations are entirely determined by $Q'(\alpha)$, and we approximate $\A(\alpha) \simeq B(z_1) Q(\alpha)$. For our ease of calculations, we can shift the $\alpha$ coordinate to $x = \alpha -z_1$ without any loss of generality. Then the approximate amplitude reads:
	 \begin{equation}
	 	\A(x) \simeq B(0) \, (x+u)^{m_0}x^{m_1}(x-v)^{m_2},\qquad B(0)\neq0.  \label{eq:multfactor}
	 \end{equation}
	 
	 \ndt With this approximation, one can now obtain the stationarity equation away from the three roots:
	\begin{equation}
		\frac{\A'(x)}{\A(x)}
		=\frac{m_0}{x+u}+\frac{m_1}{x}+\frac{m_2}{x-v}=0.           \label{eq:multlogderivative}
	\end{equation}
	 
	 \ndt Solving this, one obtains the peak locations:
	 \begin{align}
	 	x_{\pm} = \frac{ m_1 (v-u)-m_0 u+m_2 v \pm \sqrt{4 m_1 D u v+Q^2}}{2 D}
	 \end{align}
	 where, $D=(m_0+m_1+m_2), \, Q=(m_1+m_2)u-(m_0+m_1)v$. 
	 \\
	 
	 The product of the two solutions is $-m_1uv/D<0$.  Hence one lies in the left cell $(-u,0)$ and the other in the right cell $(0,v)$. These are the two positive peaks of $|\A|$ in the adjacent zero-free cells. Then the separation of the peaks is,
	 \begin{equation}
	 		R_{\boldsymbol m}(u,v)
	 		=\frac{1}{D}
	 		\sqrt{Q^2+4 D m_1uv}.                                      \label{eq:radius}
	 \end{equation}

	\subsection{Ensemble average and small spacing probability}
	
	For a given set of multiplicities $\boldsymbol m$ and the consecutive roots $(z_0, z_1, z_2)$, let $D=m_0+m_1+m_2$ be the total degree of the approximated amplitude polynomial.  As discussed earlier, the amplitude near three consecutive roots (see \eqref{eq:localfactor}) can be expanded as,
	\begin{equation}
		\A(\alpha) = c\prod_{j=0}^{2} (\alpha - z_j)^{m_j}
		= \sum_{k=0}^{D} a_k \alpha^{D-k}  . 
		\label{eq:multpoly}
	\end{equation}
	where, the $a_i$ coefficients are functions of elementary symmetric polynomials of the roots. First few of them can be expressed as,
	\begin{equation}
		a_0 = c, \;
		a_1 = c \, p_1,\;
		a_2=\frac{c}{2} \left( p_1^2-p_2 \right),\;
		a_3=\frac{c}{6} \left( p_1^3 - 3 p_1 p_2 + 2 p_3 \right); \qquad
		p_k=\sum_{j=0}^{2}m_jz_j^k.                          \label{eq:newtonthree}
	\end{equation}

	Now consider some parameters in the amplitude (for strings it is the integer partition modes) being drawn from a distribution. Then sampling the consecutive triple roots for different amplitudes from an ensemble essentially translate into choosing different sets of the coefficients $\{a_i\}$ of the polynomial in $\alpha$. We assume these parameters are being drawn from some joint probability density of these coefficients. Since we are interested in the small spacing limit of three consecutive roots, which can be described by the four parameters $(c, z_0, z_1, z_2)$, we correspondingly choose four coefficients $(a_0, a_1, a_2, a_3)$ which will be sufficient to describe the colliding limit of the three roots. Hence, we assume these relevant coefficients are being drawn from a joint distribution of the form $p(a_0, a_1,a_2,a_3,\eta) \equiv p(\ba, c, \eta)$, where $\ba = (a_1, a_2, a_3)$, and the remaining background coefficients are collectively denoted as $\eta$. Then by definition, the probability of these variables lie in an infinitesimal box,
	\begin{equation}
		\dd\Pr
		=p(\ba,c,\eta)\,
		\dd a_1\,\dd a_2\,\dd a_3\,\dd c\,\dd\eta .       \label{eq:coefficientprobabilityelement}
	\end{equation}
	
	\ndt Changing the variables from $\ba \to \bz \equiv (z_0, z_1, z_2)$ induces the Jacobian,
	\begin{equation}
		\begin{split}
			\left|\frac{\partial(a_1,a_2,a_3)}
			{\partial(z_0,z_1,z_2)}\right|
			=|c|^3| \, m_0 m_1 m_2 \, |(z_1-z_0)(z_2-z_1)(z_2-z_0)|
			= C_{\boldsymbol{m}} \, |\Delta_3(\bz)| ,               \label{eq:coefficientjacobian}
		\end{split}
	\end{equation}
	where, $C_{\boldsymbol{m}} = |c|^3| \, m_0 m_1 m_2$, and $ |\Delta_3 (\bz)| = |(z_1-z_0)(z_2-z_1)(z_2-z_0)|$ is the Vandermonde determinant like factor\footnote{This is algebraically the same Vandermonde determinant that appears in random-matrix theory, \(\Delta(z)=\prod_{i<j}(z_j-z_i)\), but the probability measures are different.  In random matrix theory, changing variables from matrix entries to eigenvalues and eigenvectors produces \(|\Delta|^{\beta_{\rm D}}\), with the Dyson index \(\beta_{\rm D}=1,2,4\) for the standard real, complex, and quaternionic symmetry classes~\cite{Forrester2010}.  Here we change variables from the three coefficients of a real cubic to its three roots, so the Jacobian contains only one power \(|\Delta_3|\).  No random-matrix ensemble or Dyson index is being assumed.  In particular, the spacing exponent \(\beta\) derived later is not \(\beta_{\rm D}\): it comes from three powers of the present Vandermonde and one power from the two-gap radial measure, provided the residual weight is regular.}. Then we have $\dd a_1\,\dd a_2\,\dd a_3 =  C_{\boldsymbol{m}} |\Delta_3| \dd z_0 \, \dd z_1 \, \dd z_2$, and hence the resulting joint density of the three roots $\bz$ and $c, \eta$ is:
	\begin{equation}
		\begin{split}
			q(\bz,c,\eta)
			=C_{\boldsymbol{m}} |\Delta_3(\bz)| \, p(\ba (\bz), c, \eta) \,  \mathbf 1_{\rm adm}.
	        \label{eq:jointrootbackgrounddensity}
		\end{split}
	\end{equation}
	Here $\mathbf 1_{\rm adm}$ imposes the conditions for admissible roots: the roots are physical, simple (i.e., each roots must be isolated and crossed normally: $\A (z_i) =0, \, \A' (z_i) \neq 0$), consecutive (i.e., after ordering $z_0 < z_1 < z_2$ there are no physical roots in the interval $(z_0, z_1)$ and $(z_1, z_2)$), and in the chosen bulk region (i.e., the three roots must lie away from endpoints, poles etc). Only when all these conditions are met, $ \mathbf 1_{\rm adm}$ is equal to unity, otherwise it is 0.
	
	Since only the information of the three roots are important for measuring the gap between two adjacent peaks (the coefficient $c$ does not change the location of the roots), it then implies that all the remaining variables $(c, \eta)$ should be integrated over to get the density of the root triples. Hence we obtain, the density of the ordered roots triples $\bz$:
	\begin{equation}
			g(\bz)
			= \int\!\dd c\,\dd\eta\; q(\bz, c, \eta) 
			= |\Delta_3(\bz)| \,  \int\!\dd c\,\dd\eta\; C_{\boldsymbol{m}}\, p(\ba (\bz), c, \eta) \,  \mathbf 1_{\rm adm} .
			\label{eq:rootdensityfromcoefficients}
	\end{equation}
	
	\ndt In the above expression, notice that the density has a prefactor of  $|\Delta_3(\bz)|$. This factor disappears if any of the roots approach the other, i.e., the probabilitty of the roots approaching each other is vanishing, and this is the origin of the peak repulsion as we will see soon explicitly. Now note, we want to obtain the behaviour of the small peak spacing distribution. From our previous discussion in Eq.\eqref{eq:radius}, we saw that the adjacent peak spacing is a function of only $u = z_1 - z_0$ and $v = z_2 - z_1$. This implies, to obtain the spacing distribution, we need to first change variables $(z_0, z_1, z_2) \to (z_1, u, v)$ and then integrate over $z_1$. Thus, the density of having three roots with gaps $u, v$ is:
	\begin{equation}
		\begin{split}
			f(u, v) &=  \int\!\dd z_1 \, |\Delta_3(\bz)| \,  \int\!\dd c\,\dd\eta\; C_{\boldsymbol{m}} \, p(\ba (\bz), c, \eta) \,  \mathbf 1_{\rm adm} \\
			&= u v (u+v)  \, \int\!\dd z_1 \,  \int\!\dd c\,\dd\eta\; C_{\boldsymbol{m}} \, p(\ba (u, v, z_1), c, \eta) \,  \mathbf 1_{\rm adm} \\
			&= u v (u+v) \, \mathcal W(u,v),
		\end{split}
	\end{equation}
	where, we have used $|\Delta_3(\bz)| = u v (u+v)$, and the Jacobian of $(z_0, z_1, z_2) \to (z_1, u, v)$ is 1.
	
	Now we are ready to define the probability density $\mP_{\rm root} (r)$ of two adjacent peaks having the gap $r$, obtained from the three consecutive roots:
	 \begin{equation}
	 	\mP_{\rm root} (r)=
	 	\int_0^\infty\!\dd u\int_0^\infty\!\dd v\;
	 	f(u,v)\,\delta\! \left(r-R_{\boldsymbol m}(u,v) \right).             \label{eq:radialintegral}
	 \end{equation}
	 
	 Since we are considering only $u, v>0$, the ordering of the roots are already maintained. To extract the leading $r$ behaviour, we define the variables $(r,t)$:
	 \begin{equation}
	 	r = R_{\boldsymbol m}(u,v) , \qquad t=\frac vu>0,
	 \end{equation}
	 which then imply,
	 \begin{equation}
	 	\begin{gathered}
	 		b_{\boldsymbol{m}}(t)=\frac{D}{\sqrt{t \left(4 D m_1+\left(m_1+m_2\right){}^2 t\right)-2 \left(m_1+m_2\right) \left(m_0+m_1\right) t+\left(m_0+m_1\right){}^2}},
	 		\\
	 		u=r\,b_{\boldsymbol{m}}(t),\quad 
	 		v=r\,t\,b_{\boldsymbol{m}}(t), \quad 	
	 		\left|\frac{\partial(u,v)}{\partial(r,t)}\right| =r\,b_{\boldsymbol{m}}(t)^2, \quad  
	 		uv(u+v)=r^3 \, t(1+t) \, b_{\boldsymbol{m}}(t)^3.
	 	\end{gathered}     
	 	\label{eq:sizeshape}
	 \end{equation}
	Substituting these into \eqref{eq:radialintegral} one obtains:
	\begin{equation}
		\mP_{\rm root} (r) = r^4  \, \int_0^\infty\!\dd t\; t(1+t) \, b_{\boldsymbol{m}} (t)^5 \, \mathcal{W} (r b_{\boldsymbol{m}} (t), r t b_{\boldsymbol{m}} (t)) = r^4 \, W (r),
	\end{equation}
	where $W(r)$ denotes the integral above. If for $r \to 0$, $W(r)$ tends to a finite non-zero value $W_0$, then one obtains,
	\begin{equation}
		\boxed{\mP_{\rm root} (r) \sim r^4 \, W_0.}                \label{eq:Wdefinition}
	\end{equation}
	
	Hence the small peak spacing repulsion is quartic, i.e., $\beta =4$. The number $4$ is therefore fixed by local three-root geometry, i.e., the analytical behaviour of the amplitude function. All the details of the underlying ensemble distribution is contained in the coefficient $W_0$, and the only assumption relevant to this calculation is that $W_0$ needs to be finite, non-zero and $r$-independent in the $r \to 0$ limit.

	\subsection{Subtleties in this description}
	
	An important subtlety this analysis presents is the distinction between the actual peak spacing $s$ obtained from the exact peak locations, and, the peak spacing $r$ obtained from the effective cubic expression involving the three roots. If the ratio between these, $T = s/r$ is close to unity in the statistical ensemble, then this three-roots description holds true. As we will see in the following sections, the median $T$ remains more than 0.99 for the HTT amplitudes, indicating that most of the events satisfy $s/r \simeq 1$.
	
	As discussed in the previous section, the analyticity and the smoothness and regularity of the underlying distribution are responsible for the quartic repulsion in the amplitude peaks. In this section, we see the same quartic behaviour appearing from the analytic structure of the amplitude assumed in Eq.\eqref{eq:radius}, and the existence of a regular finite limit of the underlying distribution in the small spacing limit (reflected in the finite non-zero value of $W_0$ and its $s$-independence).
	
	In the following sections, we will numerically compute the effective $\beta$ for string amplitudes (specifically for the HTT case), and then test our analytical calculations against it.


	\section{Numerical experiments with strings}
	\label{sec:HTTnumerics}

	\subsection{The HTT amplitude and exact peak finding}
	
	For a partition \(\lambda\vdash N\), let \(n_\ell\) denote the occupation of
	the part \(\ell\), and \(L=\sum_\ell n_\ell\).  Then up to a nonzero partition-dependent constant, the HTT amplitude between a highly excited bosonic string state at level $N$ with partition $\lambda$ and two tachyons can be expressed as a function of an angular variable $\alpha$~\cite{GrossRosenhaus2021,BianchiEtAl2022}:,
	\begin{equation}
		\A_{N,\lambda}(\alpha)=\sin^L\!\alpha
		\prod_{\ell=1}^N\prod_{k=1}^{\ell-1}
		\left(\cos\alpha-\frac{N-1}{D_N}\left(2\frac{k}{\ell}-1\right)\right)^{n_\ell}.
		\label{eq:HTTfactor}
	\end{equation}
	where, $D_N=\sqrt{N^2+2N-3}$. Then the root of this amplitude labeled by \(k/\ell\) lies at,
	\begin{equation}
		\cos\alpha_{\ell k}=\frac{N-1}{D_N}\left(2\frac{k}{\ell}-1\right).
		\label{eq:HTTroot}
	\end{equation}
	
	\ndt The solutions of the above expression for all possible $\ell, k$ provides the set of the roots $\cZ_{N,\lambda}$ for a particular partition $\lambda$ of $N$. We denote the interval between two consecutive roots as a cell.  Then inside a zero-free cell,
	\begin{equation}
		\begin{gathered}
			X(\alpha)= \partial_\alpha \log |\A(\alpha)| = \frac{\A'}\A
			=L\cot\alpha-\sin\alpha\sum_{z\in\cZ_{N,\lambda}}
			\frac{M_z}{\cos\alpha-\cos z},\\
			X'(\alpha)=-L\csc^2\alpha+
			\sum_z M_z\frac{\cos\alpha\cos z-1}{(\cos\alpha-\cos z)^2}<0.
			\label{eq:Xprime}
		\end{gathered}
	\end{equation}
	The inequality is strict because \(\cos\alpha\cos z-1\le0\), while the first term is
	strictly negative. Then in terms of $X$, the peak conditions in Eq.\eqref{eq:peakcondition} translates to:
	\begin{align}
		X(\alpha_{\mathrm{peak}}) = 0, \qquad X'(\alpha_{\mathrm{peak}}) < 0.
	\end{align}
	
	\ndt This condition, along with the identities in Eq.\eqref{eq:Xprime}, has immediate consequences:
	\begin{itemize}
		\item The logarithm of the magnitude is strictly concave inside every cell is bounded by two roots.
		
		\item \(X\) starts at \(+\infty\) at the left wall of a cell and ends at \(-\infty\) at the right wall. Hence it must therefore cross zero exactly once, and this zero crossing is the position of the peak.
		
		\item Thus every active zero-free HTT cell contains exactly one peak, and every consecutive pair of peaks has exactly one amplitude zero between them.
	\end{itemize}
	This exact cell picture eliminates grid ambiguity and provides a numerically stable algorithm:
	order the exact roots, then bracket the unique zero of \(X\) in every cell to determine the peaks. With this numerical algorithm in place, partitions are sampled uniformly from all integer partitions of $N$~\cite{Fristedt1993}.

	\subsection{Methods and relevant steps}
	
	Let $S$ be the unfolded distance between consecutive physical maxima of $|\A|$.  We define the
	contact exponent by looking at the small spacing $S$ probability density of the nearest neighbour peaks,
	\begin{equation}
		p_{\rm nn}(S)\sim C \, S^\beta, \qquad S\downarrow0.                 \label{eq:defbeta}
	\end{equation}
	Equivalently, the effective cumulative distribution function (CDF) of this spacing distribution is then,
	\begin{equation}
		F_{\rm nn}(s)=\Prob(S<s)\sim \int^s \dd S \, p_{\rm nn}(S) =  \frac{C}{\beta+1}s^{\beta+1},
		\qquad
		\beta=\frac{\dd\log F_{\rm nn}}{\dd\log s}-1.                 \label{eq:cdfbeta}
	\end{equation}
	
	The density exponent is therefore $4$ when the CDF slope is $5$. Instead of using the CDF, one could plot the histogram distribution of the sampled peaks to obtain the exponent. But near $s \to 0$, very few events occur, so the resulting histogram would be sensitive to the specifics of the bins chosen, mainly the bin widths, arbitrary bin boundaries, empty or nearly empty bins and large fluctuations in individual bins. The CDF avoids these issues, so we proceed with this route. The relevant steps for this are following,
	\begin{itemize}
		\item \textbf{Common unfolding:} We define an unfolded coordinate $\xi$ universal for all $N$, in terms of which we will perform all our computational analysis~\cite{GuhrEtAl1998,Forrester2010}. We use,
		\begin{equation}
			y=-\cos\alpha,\qquad \xi=\rho_N y.                              \label{eq:coordinate}
		\end{equation}
		All roots, all exact peaks, and all separations are transformed with this same map.  The analysis
		is restricted to $-0.5\le y\le0.5$, since this region is well separated from any endpoints of the peak distribution.  An independent pilot sample of $500$ partitions determines $\rho_N$ as the mean number of physical peaks in this interval.  Since the interval has unit $y$-length, this makes the mean peak density in $\xi$ equal to one. 
		
		\item \textbf{Shrinking contact window:} Since we will be dealing with large but finite $N$, the $s \to 0$ is not a smooth process. At very low $s$ the number of events are very small. The number of these events are strongly dependent on $N$, since with increasing $N$ one expects the events to increase for typical partitions. Our analysis in the previous sections concerns the simultaneous $N \to \infty$, $(r,s) \to 0$ limit. So we correspondingly choose our fitting window for a spacing interval where the empirical CDF has the following lower and upper limits respectively,
		\begin{equation}
			q_{\rm lo}(N)=5\times10^{-4}\left(\frac{N}{320}\right)^{-\gamma},
			\qquad
			q_{\rm hi}(N)=2\times10^{-3}\left(\frac{N}{320}\right)^{-\gamma}. \label{eq:qwindow}
		\end{equation}
		This ensures that at every $N$ the number of considered events are not too few, while also the corresponding spacing is small. Since there is no unique finite $N$ window, we check the obtained $\beta$ for various window shrinking rates: $\gamma = 1/4, 1/2, 3/4$. If the inferred \(\beta\) remains compatible for several shrinking-window rates, the conclusion can then be considered as more robust.
		
		\item \textbf{CDF fit:} Numerically, we first calculate the observables ${x_{(i)}}$ and sort the observations in an increasing order: $x_{(0)} < x_{(1)} < \cdots < x_{(M)}$. Then the resulting empirical CDF at $x_{(i)}$ is~\cite{Cunnane1978,HyndmanFan1996},
		\begin{equation}
			F_i=\frac{i+1/2}{M}.                                          \label{eq:plotpos}
		\end{equation}
		A straight line is fitted to $(\log x_{(i)},\log F_i)$ in the controlled window,
		\begin{align}
			\log F_i = a + (\beta + 1) \log x_{(i)}
		\end{align}
		The slope then directly provides $(\beta + 1)$.
	\end{itemize}
	
	\ndt With these steps in place, we can now define the concrete computational questions we will explore in the subsequent sections.

	\subsection{Questions}
	\label{sec:questions}
	
	The following are our principal questions:
	\begin{enumerate}
		
		\item Does the actual amplitude peak process has $\beta \simeq 4$ in our controlled small spacing window? This is obtained from the CDF $F^N_{\mathrm{nn}} (s)$, calculated with the actual peak spacings of the amplitude.
		
		\item In Section \ref{sec:spacing analysis}, with the direct peak-spacing analysis we had obtained,
		\begin{equation}
			\mP_\cP (s) \simeq \frac{I_N}{48} s^4,  \qquad I_N = \int_{\mathbb R}|c|^3
				p^{(N)}_J(0,0,0,c)\,\dd c,
		\end{equation}
		with the assumption that $I_N$ is independent of $s$ in the small spacing limit. Note, we explicitly introduce $N$ to emphasize that our computations are for finite $N$. If the above expression holds true, then one will obtain,
		\begin{align*}
			F^N_{nn} (s)  = \int^s \dd S \, \mP_\cP (S) \simeq \frac{I_N}{240} s^5.
		\end{align*}
		Hence one can define the following function, obtained from the empirical CDF,
		\begin{align}
			\widehat I_{N} (s) = \frac{240}{s^5} F^N_{nn} (s).
		\end{align}
		Then we ask, is $\widehat I_{N} (s)$ is approximately constant in the controlled spacing window? The variation of $\widehat I_{N} (s)$ is measured from the maximum-to-minimum ratio in the window: $\Delta \widehat I_{N} = \max (\widehat I_{N} (s)) / \min (\widehat I_{N} (s))$. If $\Delta \widehat I_{N}$ is close to unity, this indicates the $s$-independence of $I_N$.
		
		\item In Section \ref{sec:roots analysis}, the consecutive roots analysis resulted in the small spacing distribution:
		\begin{equation*}
			\mP_{\rm root} (r) \sim r^4 \, W_0,
		\end{equation*}
		along with the assumption that in the colliding root limit $r \to 0$, $W_0$ reaches a non-zero finite constant plateau.To test this, we first sample all the simple roots of the amplitude and calculate the resulting peak spacings $r = R(u,v)$ following Eq.\eqref{eq:radius}. Then we calculate $F^N_{\rm root} (r)$ using these values of $r$, and focus in our controlled window. If the function,
		\begin{align}
			\widehat W_{N} (r) = \frac{5}{r^5} F^N_{\rm root} (r)
		\end{align}
		remains approximately constant in the window, then it will indicate the $r$-independence of $W_0$. Similar to the previous step, we denote the variation of $W_N$ with the ratio: $\Delta \widehat W_{N} = \max (\widehat W_{N} (r)) / \min (\widehat W_{N} (r))$. Hence we ask, is $\Delta \widehat W_{N} $ approximately close to unity in the window?
		
		\item While discussing the analysis in Section \ref{sec:roots analysis}, we had pointed out that one should distinguish between the actual peak spacing $s$, and, the peak spacing $r$ obtained from the surrounding roots. To test this, we compute the ratio $T = s/r$ for the amplitude ensembles, and then ask: is the median $T$ close to unity? If the median is indeed very close to unity, it will indicate that statistically most of the events satisfy $s/r \simeq 1$, and the analysis worked out in Section \ref{sec:roots analysis} is indeed an equivalent description of the direct peak-spacing analysis worked out in Section \ref{sec:spacing analysis}. 
	\end{enumerate}


	\subsection{Results}
	
	For every consecutive peak pair in the amplitude, we retain the exact spacing $s$, the three bounding root locations, their multiplicities $(m_0,m_1,m_2)$, the generalized radius
	$R_{\boldsymbol m}$, and the partition identifier.  The numerical calculation includes all
	consecutive physical peaks and their event-matched generalized root triples. The number of partition samples and the number of corresponding events are noted in Table \ref{tab:counts}.
	
	\begin{table}[H]
		\centering
		\caption{Independent amplitudes and retained events.}
		\label{tab:counts}
		\begin{tabular}{rrr}
			\toprule
			$N$ & partitions & number of root-triples\\
			\midrule
			320 & 6,000 & 439,932\\
			400 & 7,000 & 637,936\\
			480 & 8,000 & 873,528\\
			560 & 9,000 & 1,143,998\\
			640 & 10,000 & 1,451,242\\
			800 & 11,000 & 1,986,992\\
			960 & 12,000 & 2,601,816\\
			1280 & 14,000 & 4,018,292\\
			\bottomrule
		\end{tabular}
	\end{table}
	
	\subsubsection{Measured contact exponent $\beta$}
	
	The measured value of $\beta$ from the amplitude ensembles for different $N$ is tabulated in Table \ref{tab:gammas}, for different various window shrinking rate $\gamma$ (see Eq.\eqref{eq:qwindow}). Here, $\beta_{\rm peak}$ denotes the exponent measured from the actual peak spacings $s$, while $\beta_{\rm root}$ denotes the exponent measured from the peak spacing $r$ derived from the corresponding root-triples.
	
	\begin{table}[H]
		\centering
		\caption{Measured exponents}
		\label{tab:gammas}
		\begin{tabular}{rcccccc}
			\toprule
			& \multicolumn{2}{c}{$\gamma=1/4$}
			& \multicolumn{2}{c}{$\gamma=1/2$}
			& \multicolumn{2}{c}{$\gamma=3/4$} \\
			\cmidrule(lr){2-3}
			\cmidrule(lr){4-5}
			\cmidrule(lr){6-7}
			$N$
			& $\beta_{\rm peak}$ & $\beta_{\rm root}$
			& $\beta_{\rm peak}$ & $\beta_{\rm root}$
			& $\beta_{\rm peak}$ & $\beta_{\rm root}$ \\
			\midrule
			320  & 4.06 & 4.02 & 4.06 & 4.02 & 4.06 & 4.02 \\
			400  & 4.04 & 4.11 & 4.06 & 4.13 & 4.07 & 4.13 \\
			480  & 4.20 & 4.02 & 4.16 & 4.00 & 4.18 & 4.02 \\
			560  & 4.03 & 3.83 & 4.06 & 3.89 & 4.08 & 4.02 \\
			640  & 4.55 & 4.43 & 4.72 & 4.49 & 4.83 & 4.53 \\
			800  & 4.16 & 4.19 & 4.22 & 4.18 & 4.38 & 4.21 \\
			960  & 4.03 & 3.93 & 4.15 & 4.09 & 4.27 & 4.21 \\
			1280 & 4.09 & 4.01 & 4.20 & 4.08 & 4.46 & 4.37 \\
			\bottomrule
		\end{tabular}
	\end{table}
	
\begin{figure}[H]
		\centering
		\includegraphics[width=0.91\linewidth]{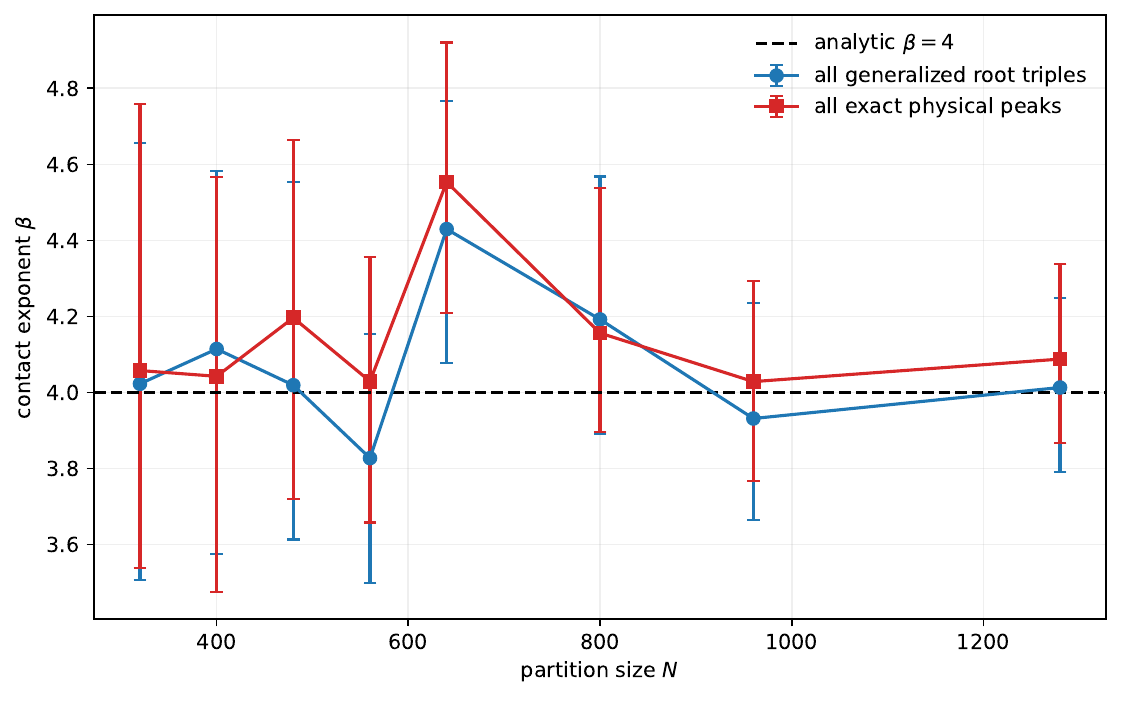}
		\caption{Primary-window exponents with whole-partition bootstrap errors~\cite{Efron1979}.  Every generalized root
			radius and exact peak spacing comes from the same event in the same unfolded coordinate.}
		\label{fig:betas}
	\end{figure}

	We see that both $\beta_{\rm peak}$ and $\beta_{\rm root}$ stay consistently near 4 for all values of $N$, see Figure \ref{fig:betas}. Since this exponent remains consistent for all $\gamma$, we can conclude the quartic repulsion is robust throughout the ensembles. Hence this answers our first question in Section \ref{sec:questions}: the HTT amplitude indeed shows quartic peak-repulsion in small spacings.
	
	\subsubsection{Numerical test of the analytics}

\begin{figure}[H]
		\centering
		\includegraphics[width=0.98\linewidth]{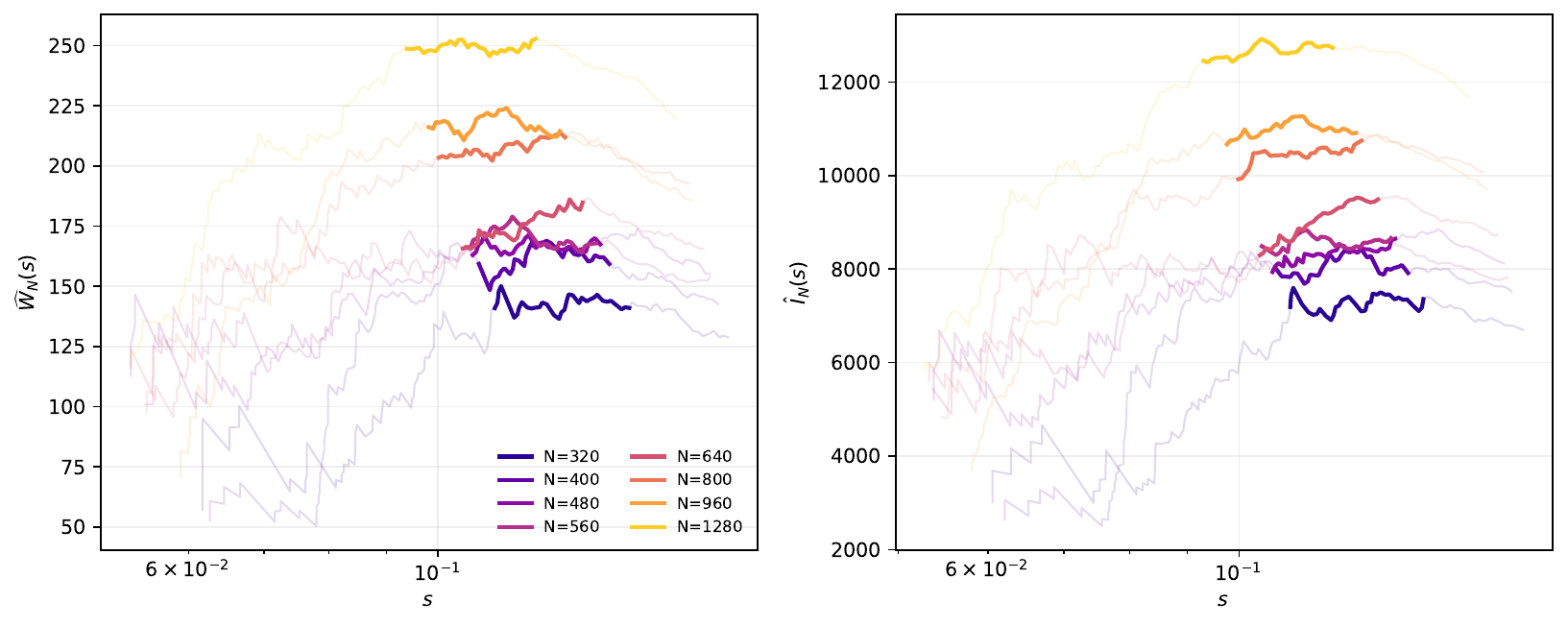}
		\caption{All-event contact estimators as functions of the common unfolded contact scale $s$.
			Left: the root-based estimator $\widehat{W}_N(s)\equiv 5F_{\rm root}(s)/s^5$.
			Right: the direct peak estimator
			$\widehat{I}_N(s)\equiv 240F_{\rm peak}(s)/s^5$; equivalently,
			$\widehat{I}_N=48\,[5F_{\rm peak}/s^5]$.  Faint curves show the diagnostic range and
			thick segments show the controlled primary windows for $\gamma=1/4$.}
		\label{fig:coefficients}
	\end{figure}

	Table~\ref{tab:coefficients} and Fig.~\ref{fig:coefficients} test the
	analytical approximations asked in the second and third points of Section \ref{sec:questions}: are $\widehat{W}_N (r)$ and $\widehat{I}_N (s)$ approximately $r, s$ independent in the controlled small spacing window? 
	
	The median $\widehat W_N$ is positive and finite, and its maximum-to-minimum variation is
	$3.5$--$16.3\%$ across the controlled windows and is only $3.5\%$ at $N=1280$. This implies that $\Delta \widehat{W}_N$ remains very close to unity in the window range, and $\widehat{W}_N (r)$ varies very little indicating the approximate $r$-independence of $W_N$. Similarly for $\widehat{I}_N$, we see that the variation is $10.2\%,10.2\%,10.1\%,7.7\%,16.3\%,9.1\%,6.2\%,5.0\%$ for increasing $N$. Hence the maximum-to-minimum variation remains between $5.0$--$16.3\%$ in the small spacing window, indicating the approximate $s$-independence of $I_N$. Both of these, then verify the validity of the analytical approximations performed in Sections \ref{sec:spacing analysis} and \ref{sec:roots analysis}, in the case of large but finite $N$ HTT amplitudes.
	
	\begin{table}[H]
		\centering
		\caption{Contact-coefficient tests in the primary windows.}
		\label{tab:coefficients}
		\begin{tabular}{rcccc}
			\toprule
			$N$ & median $\widehat{W}_N$ & $\Delta\widehat{W}_N$
			& median $\widehat{I}_N$ & $\Delta\widehat{I}_N$\\
			\midrule
			320 & 142.8 & 1.115 & 7,300 & 1.102\\
			400 & 162.5 & 1.163 & 8,061 & 1.102\\
			480 & 166.4 & 1.066 & 8,380 & 1.101\\
			560 & 167.8 & 1.098 & 8,583 & 1.077\\
			640 & 179.7 & 1.135 & 9,311 & 1.163\\
			800 & 208.9 & 1.065 & 10,502 & 1.091\\
			960 & 216.1 & 1.067 & 10,972 & 1.062\\
			1280 & 249.2 & 1.035 & 12,733 & 1.050\\
			\bottomrule
		\end{tabular}
	\end{table}
	
\begin{figure}[H]
		\centering
		\includegraphics[width=0.98\linewidth]{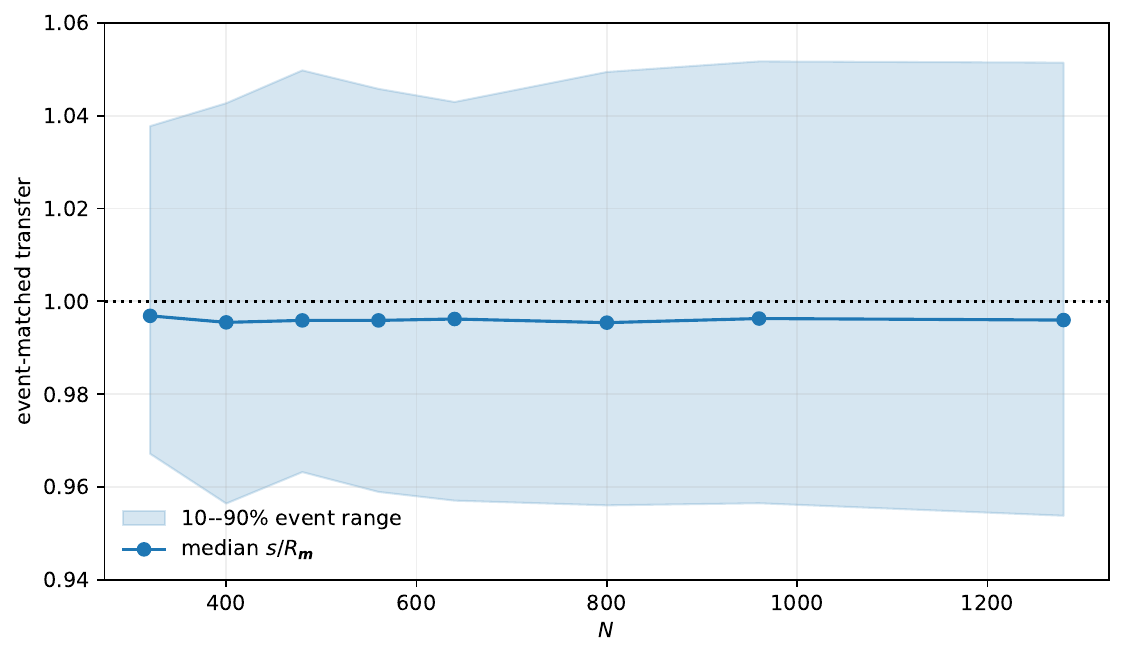}
		\caption{The median $T_{\boldsymbol{m}}$ as a function of $N$.  The shaded region
			is the event-level 10--90 percentile range.}
		\label{fig:transfer}
	\end{figure}

	For our last question of Section \ref{sec:questions}, we look at the distribution of $T_{\boldsymbol{m}} = s/ R_{\boldsymbol{m}}$, i.e., the ratio between the exact spacing $s$ and the generalized radius $R_{\boldsymbol m}$, obtained from the
	multiplicities of the three roots stored for every peak pair. The median $T_{\boldsymbol{m}}$ is noted in Table \ref{tab:median-transfer} (also plotted in Fig. \ref{fig:transfer}), which remains very close to unity ($> 0.99$) for all $N$. This directly validates the equivalency between the three-roots analysis in Section \ref{sec:roots analysis} and the direct peak-spacing analysis in Section \ref{sec:spacing analysis}.
	
	\begin{table}[H]
		\centering
		\caption{Median event-matched root-to-peak transfer factor $T_{\boldsymbol m}$.}
		\label{tab:median-transfer}
		\small
		\setlength{\tabcolsep}{5pt}
		\begin{tabular}{rcccccccc}
			\toprule
			$N$ & 320 & 400 & 480 & 560 & 640 & 800 & 960 & 1280\\
			\midrule
			median $T_{\boldsymbol m}$
			& 0.9969 & 0.9955 & 0.9959 & 0.9959 & 0.9962 & 0.9954 & 0.9963 & 0.9960\\
			\bottomrule
		\end{tabular}
	\end{table}
	
	\subsection{Conclusive summary}
	
	The numerical analysis performed in this section answers our questions regarding the HTT amplitudes conclusively. Indeed, the small peak-spacing distributions remain quartic, and our analytical explanations of this repulsion hold true approximately for the large and finite $N$ examples. Although these examples do not conclusively proof the quartic repulsion in the true $N \to \infty, \, s \to 0$ limit, they are still concrete evidences towards this asymptotic limit.
	
	Since our analytical approximations hold true for any smooth distribution of well behaved differentiable functions, the immediate next steps would be to check whether we can obtain the quartic repulsion for a different toy model examples, and whether or not the contact exponent $\beta$ changes with the change of differentiability of the amplitude functions.


	\section{Experiments and examples}
	\label{sec:examples}
	
	In this section, we will look at some particular examples, which will vary one ingredient at a time: the functions used in the amplitude, the
	coefficient law, and the local differentiability. They test the scope of the analytic arguments made in the previous sections.

	\subsection{A toy amplitude with integer partitions}

	Let $\lambda$ be an integer partition of $N$~\cite{Fristedt1993} and let $m_n$ be the multiplicity
	of the part $n$.  The toy amplitude we consider here is,
	\begin{equation}
		\A_\lambda(\alpha)
		=\sum_{n=1}^{N} m_n\sin(n\alpha),
		\qquad
		\sum_{n\geq1}n m_n=N.
		\label{eq:partition-sine}
	\end{equation}
	
	\ndt This is an amplitude dependent on integer partitions of $N$, similar to the HTT amplitudes, but the functions appearing in the amplitude are simply a linear combination of $\sin (n \alpha)$. Since this amplitude is a well-behaved analytic function, we expect the arguments of Section~\ref{sec:spacing analysis} and Section~\ref{sec:roots analysis} to hold, and the resultant small spacing behaviour should show quartic repulsion.
	 
\begin{figure}[ht]
	 	\centering
	 	\includegraphics[width=0.82\textwidth]{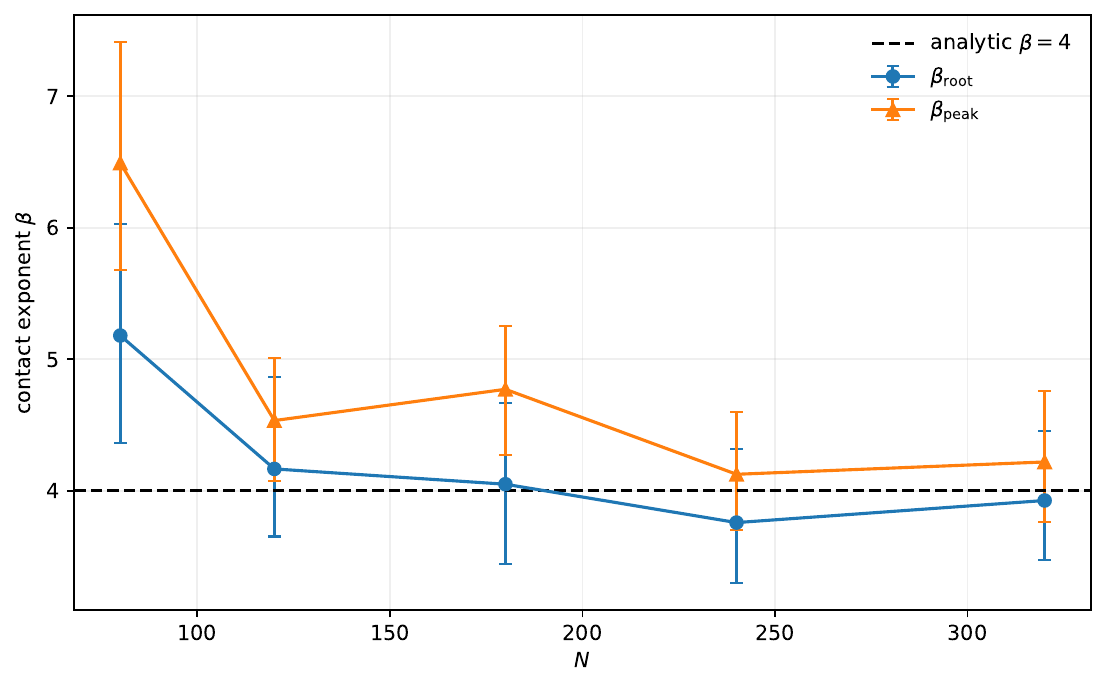}
	 	\caption{$\beta_{\rm root}$ from all-multiplicity three-root events and
	 		$\beta_{\rm peak}$ from all physical peaks in the analytic partition-sine
	 		ensemble.  The horizontal line is the analytic prediction $\beta=4$.}
	 	\label{fig:partition-sine-beta}
	 \end{figure}

	 To test this numerically, we sample $10^5$ partitions uniformly for each $N=80,120,180,240,320$. Similar to the HTT amplitudes, we consider an $N$-dependent window to look for small-$s$ behaviour of the amplitude peaks. Table~\ref{tab:partition-sine-beta} contains the contact exponents $\beta_{\rm peak}$ calculated from the direct peak spacings and  $\beta_{\rm root}$ computed from the roots surrounding the peaks; Fig.\ref{fig:partition-sine-beta} presents this visually. With increasing $N$, the small-spacing repulsion exponent closer to $N$, thus indicating quartic repulsion.

	 \begin{table}[H]
	 	\centering
	 	\small
	 	\begin{tabular}{c@{\qquad}c@{\qquad}c}
	 		\toprule
	 		$N$ & $\beta_{\rm root}$ & $\beta_{\rm peak}$\\
	 		\midrule
	 		80  & $5.18\ [4.36,6.03]$ & $6.49\ [5.68,7.41]$\\
	 		120 & $4.17\ [3.65,4.86]$ & $4.53\ [4.08,5.01]$\\
	 		180 & $4.05\ [3.44,4.66]$ & $4.77\ [4.27,5.25]$\\
	 		240 & $3.76\ [3.30,4.32]$ & $4.13\ [3.70,4.60]$\\
	 		320 & $3.93\ [3.47,4.46]$ & $4.22\ [3.76,4.76]$\\
	 		\bottomrule
	 	\end{tabular}
	 	\caption{Finite-$N$ contact exponents for the analytic partition-sine model. Numbers in brackets are partition-bootstrap 95\% intervals~\cite{Efron1979}. With increasing $N$, both $\beta_{\rm root}$ and $\beta_{\rm peak}$ become closer to 4.}
	 	\label{tab:partition-sine-beta}
	 \end{table}

	 \subsection{Examples with different distributions}
	 
	 To look into the effect of the underlying ensemble, we now consider an amplitude as a Fourier series, whose coefficients are drawn from different distributions~\cite{IksanovKabluchkoMarynych2016,DoNguyenNguyen2022}. To this end, we define:
	 \begin{equation}
	 	\A_K(\alpha)=\sum_{\ell=1}^{K}
	 	\left[a_\ell\cos(\ell\alpha)+b_\ell\sin(\ell\alpha)\right],
	 	\qquad 0\leq\alpha<2\pi.
	 	\label{eq:stationary-fourier}
	 \end{equation}
	 For each realization, all $2K$ coefficients are drawn independently from the
	 same centered distribution.  Every finite-$K$ realization is a trigonometric polynomial and hence a real analytic function~\cite{Dunnage1966,CramerLeadbetter1967,FilipJaveedTrefethen2019}.  Five different distributions of $(a_\ell, b_\ell)$ are tested:
	 \begin{center}
	 	\begin{tabular}{p{0.20\textwidth}p{0.68\textwidth}}
	 		\toprule
	 		ensemble & normalized density of each $a_\ell$ and $b_\ell$\\
	 		\midrule
	 		Gaussian & $\displaystyle
	 		f_{\rm G}(x)=\frac{1}{\sqrt{2\pi}}e^{-x^2/2}$\\[1.5ex]
	 		uniform & $\displaystyle
	 		f_{\rm U}(x)=\frac{1}{2\sqrt3}\,\1_{|x|\leq\sqrt3}$\\[1.5ex]
	 		Laplace & $\displaystyle
	 		f_{\rm L}(x)=\frac{1}{\sqrt2}e^{-\sqrt2|x|}$\\[1.5ex]
	 		Student-$t_5$ & $\displaystyle
	 		f_{t_5}(x)=\frac{8}{3\pi\sqrt3}
	 		\left(1+\frac{x^2}{3}\right)^{-3}$\\[1.5ex]
	 		smooth compact & $\displaystyle
	 		f_{\rm C}(x)=\frac{1}{Z}\exp\!\left[-\frac{1}{1-x^2}\right]
	 		\1_{|x|<1},\quad
	 		Z=\int_{-1}^{1}\exp\!\left[-\frac{1}{1-y^2}\right]\dd y$\\
	 		\bottomrule
	 	\end{tabular}
	 \end{center}
	 This collection of ensembles spans
	 light-tailed, exponential-tailed, polynomial-tailed, and compactly supported
	 continuous distributions.
	 
\begin{figure}[ht]
	 	\centering
	 	\includegraphics[width=0.88\textwidth]{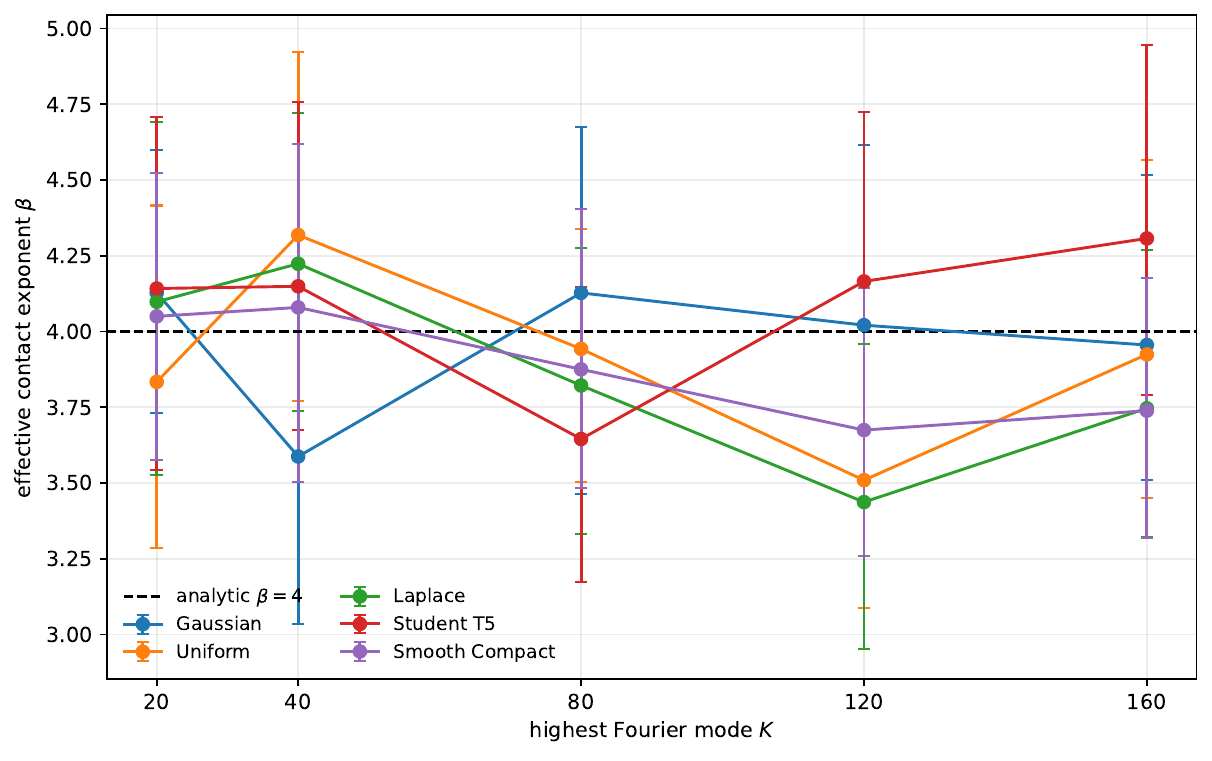}
	 	\label{fig:fourier-beta}
	 	\caption{Effective contact exponent $\beta$ for the five random
	 		Fourier ensembles as a function of the highest mode $K$.  The dashed line marks the analytic
	 		prediction $\beta=4$.}
	 \end{figure}

	 Since all of these distributions are smooth and well behaved, one would expect the analytical arguments of Sections \ref{sec:spacing analysis} and \ref{sec:roots analysis} to hold, and the resulting small spacing repulsion would be quartic. Indeed, when checked numerically for $K = 20, 40, 80, 120, 160$, the resulting $\beta$ consistently remains close to 4 for all the five ensembles, see Fig.\ref{fig:fourier-beta}.

	 \subsection{Differentiability and the contact exponent}
	 
	 To understand how differentiability affects the contact exponent $\beta$, we now consider,
	 \begin{equation}
	 	\A_\gamma(x)=a-\mu x+\operatorname{sgn}(x)|x|^\gamma,
	 	\qquad \mu>0.
	 	\label{eq:cusp-amplitude}
	 \end{equation}
	 The two random variables $a$ and $\mu$ have independent, smooth, nonzero
	 densities near the origin.  The differentiability at $x=0$ is controlled by
	 $\gamma$:
	 \begin{itemize}
	 	\item $1<\gamma\leq2$: $A_\gamma$ is $C^1$ but not $C^2$;
	 	\item $2<\gamma<3$: $A_\gamma$ is $C^2$ but not $C^3$;
	 	\item $\gamma=3$: $\operatorname{sgn}(x)|x|^3=x^3$, so the collision profile
	 	is the analytic cubic.
	 \end{itemize}
	 Here $C^k$ means that the function and its first $k$ derivatives are
	 continuous.  Thus $C^1\setminus C^2$ means ``once continuously differentiable,
	 but not twice continuously differentiable''. For $x\neq0$, we can solve for the adjacent stationary points $x_{\pm}$ to obtain,
	\begin{equation}
		x_\pm=\pm h,
		\qquad
		h=\left(\frac{\mu}{\gamma}\right)^{1/(\gamma-1)},
		\qquad
		s=x_+-x_-=2h.
		\label{eq:cusp-roots}
	\end{equation}
	Both of these stationary points are peaks of $|A_\gamma|$ precisely when,
	\begin{equation}
		A_\gamma(-h)>0,
		\quad
		A_\gamma(h)<0,
		\qquad
		\Rightarrow
		\qquad
		|a|<(\gamma-1)h^\gamma.
		\label{eq:cusp-physical-window}
	\end{equation}
	The width of the allowed $a$ interval is therefore proportional to $s^\gamma$.
	
	Let $f_a(a)$ and $f_\mu(\mu)$ be the smooth probability densities of the two
	independent random variables.  Assume that both are continuous and nonzero at
	the origin.  Before normalizing by the total probability of producing two
	physical peaks, their spacing density is
	\begin{equation}
		\mP_\gamma(s)
		=\int_0^\infty\!\dd\mu\,f_\mu(\mu)
		\int_{-\infty}^{\infty}\!\dd a\,f_a(a)\,
		\1_{\{|a|<(\gamma-1)h(\mu)^\gamma\}}
		\delta\!\left[s-2h(\mu)\right],
		\label{eq:cusp-spacing-integral}
	\end{equation}
	where $h(\mu)=(\mu/\gamma)^{1/(\gamma-1)}$.  This averages over all draws of
	$(a,\mu)$, retains the draws satisfying the physical-peak condition, and uses
	the delta function to select those whose spacing equals $s$.
	
	Changing the variable from $\mu$ to $s$ gives,
	\begin{equation*}
		\mu(s)=\gamma\left(\frac{s}{2}\right)^{\gamma-1},
		\qquad
		\left|\frac{\dd\mu}{\dd s}\right|
		=\frac{\gamma(\gamma-1)}{2}
		\left(\frac{s}{2}\right)^{\gamma-2},
		\qquad
		\delta[s-2h(\mu)]
		=\delta[\mu-\mu(s)]|\dd\mu/\dd s|,
	\end{equation*}
	and hence one obtains,
	\begin{equation}
		\mP_\gamma(s)
		=f_\mu\!\left(\gamma(s/2)^{\gamma-1}\right)
		\frac{\gamma(\gamma-1)}{2}
		\left(\frac{s}{2}\right)^{\gamma-2}
		\int_{-(\gamma-1)(s/2)^\gamma}^{+(\gamma-1)(s/2)^\gamma}
		f_a(a)\dd a.
		\label{eq:cusp-exact-density}
	\end{equation}

	\ndt Since we are considering small spacing behaviour, we can approximate the two densities smoothly as,
	\begin{align*}
		f_\mu\!\left(\gamma(s/2)^{\gamma-1}\right)
		&=f_\mu(0)+o(1),\\
		\int_{-(\gamma-1)(s/2)^\gamma}^{+(\gamma-1)(s/2)^\gamma}
		f_a(a)\dd a
		&=2(\gamma-1)f_a(0)\left(\frac{s}{2}\right)^\gamma
		+o(s^\gamma),
	\end{align*}
	and therefore,
	\begin{align}
		\mP_\gamma(s)
		& \simeq \gamma(\gamma-1)^2f_a(0)f_\mu(0)
		\left(\frac{s}{2}\right)^{2\gamma-2}.
		\label{eq:cusp-leading-density}
	\end{align}
	
	\ndt Hence the normalized physical
	spacing density has a $\gamma$ dependent contact exponent:
	\begin{equation}
		\boxed{\beta(\gamma)=2\gamma-2.}
		\label{eq:tunable-beta}
	\end{equation}
	This gives a continuous interpolation from weak repulsion to the analytic
	quartic law.
	
\begin{figure}[ht]
		\centering
		\includegraphics[width=\textwidth]{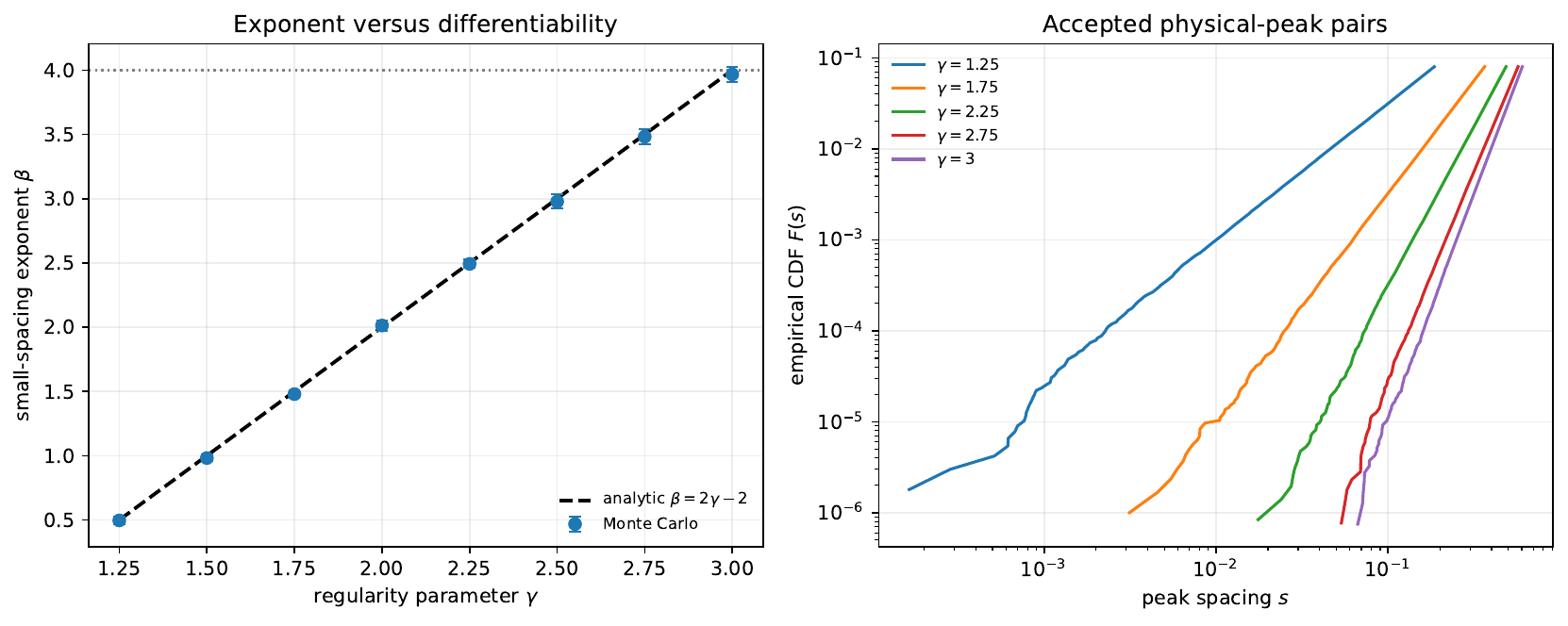}
		\caption{Left: the measured exponents lie on the exact line
			$\beta=2\gamma-2$.  Right: empirical spacing CDFs steepen continuously as the
			collision profile becomes more differentiable.}
		\label{fig:cusp-verification}
	\end{figure}

	We now test this $\gamma$ dependence of $\beta$ numerically. For each $\gamma$, $5\times10^6$ independent pairs were drawn from the uniform distributions:
	\begin{equation}
		\begin{aligned}
			\mu&\sim \operatorname{Unif}\left(0,\gamma H^{\gamma-1}\right),\\
			a&\sim \operatorname{Unif}\left(-(\gamma-1)H^\gamma,
			+(\gamma-1)H^\gamma\right),
			\qquad H=\frac12.
		\end{aligned}
	\end{equation}
	This particular scaling of the distributions with $\gamma$ ensures that every sampled amplitude has two stationary points inside the fixed interval $-1/2 \leq x \leq 1/2$, for every $\gamma$. The exact $h$ from Eq.~\eqref{eq:cusp-roots} was evaluated and the event was
	retained only when Eq.~\eqref{eq:cusp-physical-window} held. The primary CDF fit used
	the $0.2\%$--$2\%$ accepted-spacing window. Table \ref{tab:cusp-results} contains the resulting numerically obtained $\beta$ for $1.25 \leq \gamma \leq 3$. The numerical values match the analytical approximations closely, which is visually represented in Fig.\ref{fig:cusp-verification}.
	
	\begin{table}[ht]
		\centering
		\small
		\begin{tabular}{c@{\qquad}c@{\qquad}c@{\qquad}c}
			\toprule
			$\gamma$ & analytic $\beta = 2\gamma-2$ & numerical $\beta$ & 95\% interval\\
			\midrule
			1.25 & 0.50 & 0.496 & $[0.467,0.525]$\\
			1.50 & 1.00 & 0.982 & $[0.956,1.007]$\\
			1.75 & 1.50 & 1.479 & $[1.450,1.508]$\\
			2.00 & 2.00 & 2.013 & $[1.971,2.055]$\\
			2.25 & 2.50 & 2.492 & $[2.459,2.526]$\\
			2.50 & 3.00 & 2.977 & $[2.922,3.032]$\\
			2.75 & 3.50 & 3.483 & $[3.425,3.540]$\\
			3.00 & 4.00 & 3.967 & $[3.907,4.027]$\\
			\bottomrule
		\end{tabular}
		\caption{Analytic and Monte Carlo exponents for the isolated tunable cusp.
			The numerical values track $2\gamma-2$ throughout the entire interpolation.}
		\label{tab:cusp-results}
	\end{table}
	
	\subsection{Implications}
	
	The three kind of examples explored in this section has several implications. The integer partitioned analytic sine model has the same underlying distribution as the HTT amplitudes, but the constituent functions of the amplitude are different, and yet they result into quartic repulsion. And our experiments with different ensembles of the Fourier series, on the other hand, shows numerical results consistent with quartic repulsion. Then, our last set of examples with tunable differentiability show that the nature of the peak repulsion is essentially tied to the analytic behaviour of the amplitude itself. These together indicate towards the fact that, for an ensemble of amplitude to have peak-repulsion, one does not need a set of specific functions or a finely tuned ensemble distribution. If one has an ensemble of well-behaved differentiable amplitude drawn from a smooth distribution, it will exhibit quartic repulsion.


	\section{Summary and future directions}

	We began with a local question: why are two neighboring peaks of a highly excited string
	amplitude unlikely to be arbitrarily close?  The answer is not that the peak positions are
	eigenvalues of a hidden random matrix.  Two nearby physical peaks require a constrained local
	configuration of the amplitude and its derivatives.  A direct peak spacing calculation translates
	those constraints into a contact density proportional to the fourth power of the spacing.  The
	equivalent three-root description exposes the same power through a root-to-coefficient
	Vandermonde factor and the two-dimensional phase space of the adjacent root gaps.  Root
	multiplicities modify the finite prefactor but not this leading power.

	The conclusion is that for a sufficiently smooth real-function ensemble,
	quartic repulsion follows when the relevant coefficient does not vanish or blow up in the colliding limit.  The HTT
	numerics support these assumptions in the finite-
	\(N\) contact windows that contain adequate statistics.  The
	partition-sine and random-Fourier examples show that the result is insensitive to the detailed
	coefficient law within the tested smooth ensembles.  Conversely, the cusp family demonstrates
	that weakening local differentiability can change the contact power.  Several directions can now sharpen and extend this picture.
	\begin{enumerate}
		\item \textbf{A controlled large-level limit.}  The remaining mathematical step for HTT is a
		local limit theorem for the finite collection of amplitude derivatives after conditioning the
		occupation numbers on a partition of \(N\).  Such a result would turn the numerically observed
		bounded contact coefficients into a uniform statement about the order of limits
		\(N\to\infty\) and \(s\to0\).  Probabilistic descriptions of large random partitions provide a
		natural starting point~\cite{Fristedt1993}.

		\item \textbf{Beyond contact.}  The present argument fixes only the first small-spacing power.
		The bulk and tail of the spacing distribution, spacing ratios, number variance, and the
		scattering form factor probe longer-range correlations and need not be determined by the same
		local mechanism.  Existing studies already find ensemble- and kinematics-dependent behavior,
		including non-Gaussian-ensemble HTT statistics, approximately GOE-like HHT sectors, and
		mixed regular--chaotic regimes
		~\cite{BianchiEtAl2022,BianchiEtAl2023,BianchiEtAl2024,DasMandalSarkar2024,SavicCubrovic2024,BianchiEtAl2026}.
		A useful next test is to measure all of these observables on the same samples and with the same
		unfolding prescription.

		\item \textbf{More string amplitudes and kinematics.}  The analysis should be repeated for
		amplitudes with two or more highly excited external strings, closed-string states, and
		higher-point processes.  Varying polarizations, momenta, helicity sectors, and the measure on
		microstates can reveal which features are consequences of local smoothness and which retain
		genuinely string-specific information
		~\cite{GrossRosenhaus2021,Rosenhaus2022,BianchiEtAl2023,DasMandalSarkar2024,FirrottaKiritsisNiarchos2024,SavicCubrovic2024,Pesando2026}.

		\item \textbf{Many-particle QFT amplitudes.}  Erratic dependence of high-multiplicity
		\(m\to n\) amplitudes has been proposed as a scattering diagnostic of quantum chaos
		~\cite{Rosenhaus2021}.  Peak-contact statistics provide a concrete local observable for this
		proposal.  One can test whether smooth families of QFT amplitudes exhibit the same quartic law
		and then ask which additional, nonlocal observables distinguish integrable, mixed, and chaotic
		dynamics.

		\item \textbf{Loops, thresholds, and singular kinematics.}  Tree-level amplitudes are analytic
		inside the angular intervals used here.  Loop corrections introduce thresholds, branch cuts,
		and Landau singularities~\cite{Landau1959,EdenEtAl1966}.  Studying peak collisions as a
		kinematic scan approaches such a singular set would test precisely where the smooth-jet
		argument fails and whether new contact exponents or logarithmic corrections appear.

		\item \textbf{Complex amplitudes and higher-dimensional scans.}  Above threshold an amplitude
		is generally complex, and a physical peak of \(|\A|\) is no longer classified by the sign of a
		real function.  Extending Kac--Rice theory to the joint real and imaginary parts, or to peaks on
		a multidimensional kinematic manifold, should produce different codimension counting and may
		therefore define new universality classes~\cite{AdlerTaylor2007,AzaisWschebor2009,BianchiEtAl2026}.

		\item \textbf{Robustness and exceptional ensembles.}  Random trigonometric polynomials offer
		a controlled laboratory in which coefficient correlations, heavy tails, symmetry constraints,
		or discrete coefficient laws can be varied independently~\cite{Dunnage1966,IksanovKabluchkoMarynych2016,DoNguyenNguyen2022}.  These tests can
		identify the minimal regularity and nondegeneracy assumptions behind a finite contact
		coefficient, and can expose exceptions caused by singular jet measures or exact algebraic
		constraints.
	\end{enumerate}

    \begingroup
\footnotesize
\raggedright

\endgroup


\begin{thebibliography}{99}
\setlength{\itemsep}{0pt}

\bibitem{Rosenhaus2021}
V.~Rosenhaus, \emph{Chaos in the quantum field theory \(S\)-matrix},
\emph{Phys. Rev. Lett.} \textbf{127} (2021) 021601,
\href{https://doi.org/10.1103/PhysRevLett.127.021601}{doi:10.1103/PhysRevLett.127.021601}.

\bibitem{OttTel1993}
E.~Ott and T.~T\'el, \emph{Chaotic scattering: an introduction},
\emph{Chaos} \textbf{3} (1993) 417--426,
\href{https://doi.org/10.1063/1.165979}{doi:10.1063/1.165979}.

\bibitem{FyodorovSommers1997}
Y.~V.~Fyodorov and H.-J.~Sommers,
\emph{Statistics of resonance poles, phase shifts and time delays in quantum chaotic scattering: random matrix approach for systems with broken time-reversal invariance},
\emph{J. Math. Phys.} \textbf{38} (1997) 1918--1981,
\href{https://doi.org/10.1063/1.531919}{doi:10.1063/1.531919}.

\bibitem{BohigasGiannoniSchmit1984}
O.~Bohigas, M.-J.~Giannoni, and C.~Schmit,
\emph{Characterization of chaotic quantum spectra and universality of level fluctuation laws},
\emph{Phys. Rev. Lett.} \textbf{52} (1984) 1--4,
\href{https://doi.org/10.1103/PhysRevLett.52.1}{doi:10.1103/PhysRevLett.52.1}.

\bibitem{GuhrEtAl1998}
T.~Guhr, A.~M\"uller--Groeling, and H.~A.~Weidenm\"uller,
\emph{Random-matrix theories in quantum physics: common concepts},
\emph{Phys. Rep.} \textbf{299} (1998) 189--425,
\href{https://doi.org/10.1016/S0370-1573(97)00088-4}{doi:10.1016/S0370-1573(97)00088-4}.

\bibitem{Haake2018}
F.~Haake, S.~Gnutzmann, and M.~Ku\'s,
\emph{Quantum Signatures of Chaos}, 4th ed. (Springer, Cham, 2018),
\href{https://doi.org/10.1007/978-3-319-97580-1}{doi:10.1007/978-3-319-97580-1}.

\bibitem{Smilansky1991}
U.~Smilansky, \emph{The classical and quantum theory of chaotic scattering},
in \emph{Chaos and Quantum Physics}, Les Houches Session LII,
M.-J.~Giannoni, A.~Voros, and J.~Zinn-Justin, eds.
(North-Holland, Amsterdam, 1991), pp.~371--441.

\bibitem{BluemelSmilansky1988}
R.~Bl\"umel and U.~Smilansky,
\emph{Classical irregular scattering and its quantum-mechanical implications},
\emph{Phys. Rev. Lett.} \textbf{60} (1988) 477--480,
\href{https://doi.org/10.1103/PhysRevLett.60.477}{doi:10.1103/PhysRevLett.60.477}.

\bibitem{BluemelSmilansky1990}
R.~Bl\"umel and U.~Smilansky,
\emph{Random-matrix description of chaotic scattering: semiclassical approach},
\emph{Phys. Rev. Lett.} \textbf{64} (1990) 241--244,
\href{https://doi.org/10.1103/PhysRevLett.64.241}{doi:10.1103/PhysRevLett.64.241}.

\bibitem{DoronSmilansky1990}
E.~Doron, U.~Smilansky, and A.~Frenkel,
\emph{Experimental demonstration of chaotic scattering of microwaves},
\emph{Phys. Rev. Lett.} \textbf{65} (1990) 3072--3075,
\href{https://doi.org/10.1103/PhysRevLett.65.3072}{doi:10.1103/PhysRevLett.65.3072}.

\bibitem{KudrolliEtAl1994}
A.~Kudrolli, S.~Sridhar, A.~Pandey, and R.~Ramaswamy,
\emph{Signatures of chaos in quantum billiards: microwave experiments},
\emph{Phys. Rev. E} \textbf{49} (1994) R11--R14,
\href{https://doi.org/10.1103/PhysRevE.49.R11}{doi:10.1103/PhysRevE.49.R11}.

\bibitem{StoffregenEtAl1995}
U.~Stoffregen, J.~Stein, H.-J.~St\"ockmann, M.~Ku\'s, and F.~Haake,
\emph{Microwave billiards with broken time-reversal symmetry},
\emph{Phys. Rev. Lett.} \textbf{74} (1995) 2666--2669,
\href{https://doi.org/10.1103/PhysRevLett.74.2666}{doi:10.1103/PhysRevLett.74.2666}.

\bibitem{HemmadyEtAl2005}
S.~Hemmady, X.~Zheng, T.~M.~Antonsen Jr., E.~Ott, and S.~M.~Anlage,
\emph{Universal statistics of the scattering coefficient of chaotic microwave cavities},
\emph{Phys. Rev. E} \textbf{71} (2005) 056215,
\href{https://doi.org/10.1103/PhysRevE.71.056215}{doi:10.1103/PhysRevE.71.056215}.

\bibitem{MitchellEtAl2010}
G.~E.~Mitchell, A.~Richter, and H.~A.~Weidenm\"uller,
\emph{Random matrices and chaos in nuclear physics: nuclear reactions},
\emph{Rev. Mod. Phys.} \textbf{82} (2010) 2845--2901,
\href{https://doi.org/10.1103/RevModPhys.82.2845}{doi:10.1103/RevModPhys.82.2845}.

\bibitem{GrossRosenhaus2021}
D.~J.~Gross and V.~Rosenhaus, \emph{Chaotic scattering of highly excited strings},
\emph{JHEP} \textbf{05} (2021) 048,
\href{https://doi.org/10.1007/JHEP05(2021)048}{doi:10.1007/JHEP05(2021)048}.

\bibitem{BianchiEtAl2022}
M.~Bianchi, M.~Firrotta, J.~Sonnenschein, and D.~Weissman,
\emph{A measure for chaotic scattering amplitudes},
\emph{Phys. Rev. Lett.} \textbf{129} (2022) 261601,
\href{https://doi.org/10.1103/PhysRevLett.129.261601}{doi:10.1103/PhysRevLett.129.261601}.

\bibitem{BianchiEtAl2023}
M.~Bianchi, M.~Firrotta, J.~Sonnenschein, and D.~Weissman,
\emph{Measuring chaos in string scattering processes},
\emph{Phys. Rev. D} \textbf{108} (2023) 066006,
\href{https://doi.org/10.1103/PhysRevD.108.066006}{doi:10.1103/PhysRevD.108.066006}.

\bibitem{Rosenhaus2022}
V.~Rosenhaus, \emph{Chaos in a many-string scattering amplitude},
\emph{Phys. Rev. Lett.} \textbf{129} (2022) 031601,
\href{https://doi.org/10.1103/PhysRevLett.129.031601}{doi:10.1103/PhysRevLett.129.031601}.

\bibitem{HashimotoEtAl2022}
K.~Hashimoto, Y.~Matsuo, and T.~Yoda,
\emph{Transient chaos analysis of string scattering},
\emph{JHEP} \textbf{11} (2022) 147,
\href{https://doi.org/10.1007/JHEP11(2022)147}{doi:10.1007/JHEP11(2022)147}.

\bibitem{Firrotta2023}
M.~Firrotta, \emph{The chaotic emergence of thermalization in highly excited string decays},
\emph{JHEP} \textbf{04} (2023) 052,
\href{https://doi.org/10.1007/JHEP04(2023)052}{doi:10.1007/JHEP04(2023)052}.

\bibitem{SavicCubrovic2024}
N.~Savi\'c and M.~\v{C}ubrovi\'c,
\emph{Weak chaos and mixed dynamics in the string \(S\)-matrix},
\emph{JHEP} \textbf{03} (2024) 101,
\href{https://doi.org/10.1007/JHEP03(2024)101}{doi:10.1007/JHEP03(2024)101}.

\bibitem{BianchiEtAl2024}
M.~Bianchi, M.~Firrotta, J.~Sonnenschein, and D.~Weissman,
\emph{From spectral to scattering form factor},
\emph{JHEP} \textbf{06} (2024) 189,
\href{https://doi.org/10.1007/JHEP06(2024)189}{doi:10.1007/JHEP06(2024)189}.

\bibitem{BianchiEtAl2026}
M.~Bianchi, M.~Firrotta, J.~Sonnenschein, and D.~Weissman,
\emph{Multi-dimensional chaos II: string scattering amplitudes, curve repulsion, and RMT},
arXiv:2606.24490 [hep-th] (2026),
\href{https://arxiv.org/abs/2606.24490}{arXiv:2606.24490}.

\bibitem{FirrottaKiritsisNiarchos2024}
M.~Firrotta, E.~Kiritsis, and V.~Niarchos,
\emph{Scattering, absorption and emission of highly excited strings},
arXiv:2407.16476 [hep-th] (2024),
\href{https://arxiv.org/abs/2407.16476}{arXiv:2407.16476}.

\bibitem{Pesando2026}
I.~Pesando, \emph{Bosonic string massive scalar amplitudes and absence of ``chaos''},
\emph{Eur. Phys. J. C} \textbf{86} (2026) 712,
\href{https://doi.org/10.1140/epjc/s10052-026-15698-w}{doi:10.1140/epjc/s10052-026-15698-w}.

\bibitem{DasMandalSarkar2024}
D.~Das, S.~Mandal, and A.~Sarkar,
\emph{Chaotic and thermal aspects in the highly excited string \(S\)-matrix},
\emph{JHEP} \textbf{08} (2024) 200,
\href{https://doi.org/10.1007/JHEP08(2024)200}{doi:10.1007/JHEP08(2024)200}.

\bibitem{Kac1943}
M.~Kac, \emph{On the average number of real roots of a random algebraic equation},
\emph{Bull. Amer. Math. Soc.} \textbf{49} (1943) 314--320,
\href{https://doi.org/10.1090/S0002-9904-1943-07912-8}{doi:10.1090/S0002-9904-1943-07912-8}.

\bibitem{Rice1944}
S.~O.~Rice, \emph{Mathematical analysis of random noise},
\emph{Bell Syst. Tech. J.} \textbf{23} (1944) 282--332,
\href{https://doi.org/10.1002/j.1538-7305.1944.tb00874.x}{doi:10.1002/j.1538-7305.1944.tb00874.x}.

\bibitem{AzaisWschebor2009}
J.-M.~Aza\"is and M.~Wschebor,
\emph{Level Sets and Extrema of Random Processes and Fields}
(Wiley, Hoboken, 2009),
\href{https://doi.org/10.1002/9780470434642}{doi:10.1002/9780470434642}.

\bibitem{CramerLeadbetter1967}
H.~Cram\'er and M.~R.~Leadbetter,
\emph{Stationary and Related Stochastic Processes: Sample Function Properties and Their Applications}
(Wiley, New York, 1967).

\bibitem{AdlerTaylor2007}
R.~J.~Adler and J.~E.~Taylor,
\emph{Random Fields and Geometry} (Springer, New York, 2007),
\href{https://doi.org/10.1007/978-0-387-48116-6}{doi:10.1007/978-0-387-48116-6}.

\bibitem{Dunnage1966}
J.~E.~A.~Dunnage, \emph{The number of real zeros of a random trigonometric polynomial},
\emph{Proc. London Math. Soc.} \textbf{s3-16} (1966) 53--84,
\href{https://doi.org/10.1112/plms/s3-16.1.53}{doi:10.1112/plms/s3-16.1.53}.

\bibitem{FilipJaveedTrefethen2019}
S.~Filip, A.~Javeed, and L.~N.~Trefethen,
\emph{Smooth random functions, random ODEs, and Gaussian processes},
\emph{SIAM Rev.} \textbf{61} (2019) 185--205,
\href{https://doi.org/10.1137/17M1161853}{doi:10.1137/17M1161853}.

\bibitem{IksanovKabluchkoMarynych2016}
A.~Iksanov, Z.~Kabluchko, and A.~Marynych,
\emph{Local universality for real roots of random trigonometric polynomials},
\emph{Electron. J. Probab.} \textbf{21} (2016), paper no.~63, 19 pp.,
\href{https://doi.org/10.1214/16-EJP9}{doi:10.1214/16-EJP9}.

\bibitem{DoNguyenNguyen2022}
Y.~Do, H.~H.~Nguyen, and O.~Nguyen,
\emph{Random trigonometric polynomials: universality and non-universality of the variance for the number of real roots},
\emph{Ann. Inst. Henri Poincar\'e Probab. Stat.} \textbf{58} (2022) 1460--1504,
\href{https://doi.org/10.1214/21-AIHP1206}{doi:10.1214/21-AIHP1206}.

\bibitem{Forrester2010}
P.~J.~Forrester, \emph{Log-Gases and Random Matrices}
(Princeton University Press, Princeton, 2010),
\href{https://doi.org/10.1515/9781400835416}{doi:10.1515/9781400835416}.

\bibitem{Fristedt1993}
B.~Fristedt, \emph{The structure of random partitions of large integers},
\emph{Trans. Amer. Math. Soc.} \textbf{337} (1993) 703--735,
\href{https://doi.org/10.1090/S0002-9947-1993-1094553-1}{doi:10.1090/S0002-9947-1993-1094553-1}.

\bibitem{HyndmanFan1996}
R.~J.~Hyndman and Y.~Fan, \emph{Sample quantiles in statistical packages},
\emph{Amer. Statist.} \textbf{50} (1996) 361--365,
\href{https://doi.org/10.1080/00031305.1996.10473566}{doi:10.1080/00031305.1996.10473566}.

\bibitem{Cunnane1978}
C.~Cunnane, \emph{Unbiased plotting positions---a review},
\emph{J. Hydrol.} \textbf{37} (1978) 205--222,
\href{https://doi.org/10.1016/0022-1694(78)90017-3}{doi:10.1016/0022-1694(78)90017-3}.

\bibitem{Efron1979}
B.~Efron, \emph{Bootstrap methods: another look at the jackknife},
\emph{Ann. Statist.} \textbf{7} (1979) 1--26,
\href{https://doi.org/10.1214/aos/1176344552}{doi:10.1214/aos/1176344552}.

\bibitem{Landau1959}
L.~D.~Landau, \emph{On analytic properties of vertex parts in quantum field theory},
\emph{Nucl. Phys.} \textbf{13} (1959) 181--192,
\href{https://doi.org/10.1016/0029-5582(59)90154-3}{doi:10.1016/0029-5582(59)90154-3}.

\bibitem{EdenEtAl1966}
R.~J.~Eden, P.~V.~Landshoff, D.~I.~Olive, and J.~C.~Polkinghorne,
\emph{The Analytic \(S\)-Matrix} (Cambridge University Press, Cambridge, 1966).

\end{thebibliography}
\end{document}